\documentclass[amsmath,amssymb,preprintnumbers,nofootinbib,a4paper,11pt]{article}

\usepackage{graphicx} 
\usepackage{float}
\usepackage{amsmath}
\usepackage{epstopdf}
\usepackage{jheppub}
\usepackage{xspace}
\usepackage{booktabs}
\graphicspath{{./figs/}}
\usepackage{array,tabularx,epsfig,mathrsfs,graphicx,rotating}
\usepackage{ifthen}
\usepackage{amsfonts}
\usepackage{amssymb}
\usepackage{bbold}

\newcommand{\be}{\begin{equation}}
\newcommand{\ee}{\end{equation}}
\newcommand{\bea}{\begin{eqnarray}}
\newcommand{\eea}{\end{eqnarray}}

\title{
Lepton number violating operators with standard model gauge fields: 
A survey of neutrino masses from 3-loops and their link to dark matter
}

\author[1]{Michael Gustafsson}
\author[2]{, Jos\'e Miguel No}
\author[3]{, Maximiliano A. Rivera}
\affiliation[1]{Institute for theoretical Physics - Faculty of Physics, Georg-August University G\"ottingen, 
Friedrich-Hund-Platz 1, D-37077 G\"ottingen, Germany}
\affiliation[2]{Departamento de F\'isica Te\'orica and Instituto de F\'isica Te\'orica, IFT-UAM/CSIC,
Cantoblanco, 28049, Madrid, Spain}
\affiliation[3]{Departamento de F\'isica, Universidad T\'ecnica Federico Santa Mar\'ia, Casilla 110-V, 
Valparaiso, Chile}

\abstract{
We investigate neutrino mass generation scenarios where the lepton number breaking new physics couples only to the Standard Model (SM) right-handed charged lepton chirality. The lowest-order lepton number violating effective operator which describes this framework is a unique dimension nine operator involving SM gauge fields, $\mathcal{O}_9$. 
We find that there are two possible classes of new physics scenarios giving rise to this $\mathcal{O}_9$ operator. 
In these scenarios neutrino masses are induced radiatively via dark matter interactions, linking the dark matter to a natural explanation for the smallness of neutrino masses compared to the electroweak scale. 
We discuss the phenomenology and existing constraints in the different neutrino mass models within each class. 
In particular, we analyze the important interplay between neutrino mixing and neutrinoless double $\beta$-decay in order to predict characteristic signatures and disfavour certain scenarios.
}

\begin{document}


\maketitle


\section{Introduction} 

The quest to unravel the mechanism responsible for neutrino masses is a central aspect of neutrino physics, tightly linked to the Dirac or Majorana nature of neutrinos. 
Under the assumption that neutrinos are Majorana particles, a general classification of the possible lepton-number-violating higher-dimensional operators up to dimension $D = 11$ that could cause neutrino mass generation has been carried out in~\cite{Babu:2001ex,deGouvea:2007qla} (see also~\cite{Deppisch:2017ecm} for a recent discussion in the context of neutrinoless double $\beta$-decay).
This classification, aimed at providing a general understanding of new physics breaking lepton number by two units ($\Delta L = 2$), focused on higher-dimensional operators built from Standard Model (SM) quarks, leptons as well as the SM Higgs doublet.
This classification has however left-out an entire class of $\Delta L = 2$ operators, namely those containing also SM gauge fields, as these operators were initially deemed unable to accommodate a suitable renormalizable completion yielding neutrino masses.

In recent years, counterexamples to the latter argument have been found~\cite{Chen:2006vn,delAguila:2011gr,delAguila:2012nu, Gustafsson:2012vj,Alcaide:2017xoe}\footnote{See also \cite{Angel:2012ug} for a brief discussion on such counterexamples triggered by correspondence with the Authors of the present manuscript.} (see also~\cite{Cai:2017jrq} for a recent overview of neutrino mass models which include these counterexamples). 
These examples have in turn unravelled a very interesting connection between the presence of SM gauge field in the higher-dimensional $\Delta L = 2$ operator, the chirality of the SM leptons present in the operator and its lowest possible dimension $D$~\cite{delAguila:2012nu}. 
In a certain way these cases yield generalizations of the ``uniqueness'' of the Weinberg operator~\cite{Weinberg:1979sa}, which contains two left-handed chirality leptons, to alternative lowest-order $\Delta L = 2$ operators that instead have left-right (LR) or right-right (RR) chirality assignments to the SM leptons in the operator, as discussed in~\cite{delAguila:2012nu}.
In particular, for lepton number violation involving only SM leptons of right-handed chirality (the RR assignment) the lowest order $\Delta L = 2$ effective SM operator is unique and appears at $D = 9$~\cite{delAguila:2012nu,Gustafsson:2014vpa}, which in this work we label as operator $\mathcal{O}_9$. Among the salient features of $\mathcal{O}_9$ is the fact that neutrino masses from this lepton number violating operator are generated at the earliest at the 2-loop order, thereby providing a natural explanation for the smallness of neutrino masses.

In this work, we analyse in detail the structure and properties of possible renormalizable completions to the operator $\mathcal{O}_9$ in beyond Standard Model (BSM) scenarios.
At the same time, we show that for a wide range of renormalizable completions there is a connection between neutrino masses and dark matter (DM), with the DM candidate taking an active part in the mechanism of neutrino mass generation. In such cases, neutrino masses are generated at 3-loops or higher.\footnote{See~\cite{Cepedello:2018rfh} instead for an analysis of 3-loop neutrino mass models from the Weinberg operator.} We identify two general classes for renormalizable completions of the $\mathcal{O}_9$ operator, discuss the general phenomenological aspects of both classes and provide several examples of specific radiative neutrino mass models (several of them genuinely new) belonging to each class. 

\medskip
The paper is organized as follows: Section~\ref{sec:EFT} introduces the unique (dimension $D=9$) $\Delta L = 2$ RR operator $\mathcal{O}_9$, and discusses various aspects of it. We then proceed to show how every renormalizable completion of this operator can be categorised into two distinct classes of models. In section~\ref{sec:UV_complete} the two classes' distinct phenomenological features are described and several explicit model realizations are presented. The interplay of neutrino mixing and neutrinoless double $\beta$-decay within each class is studied and its phenomenological impact is discussed in section~\ref{sec:Constraints}, before we conclude in section \ref{sec:Conclusions}.

\section{Decomposing the lepton number breaking operator $\mathcal{O}_9$} 
\label{sec:EFT}

\begin{figure}[t]
\center{\includegraphics[width=.33 \textwidth]{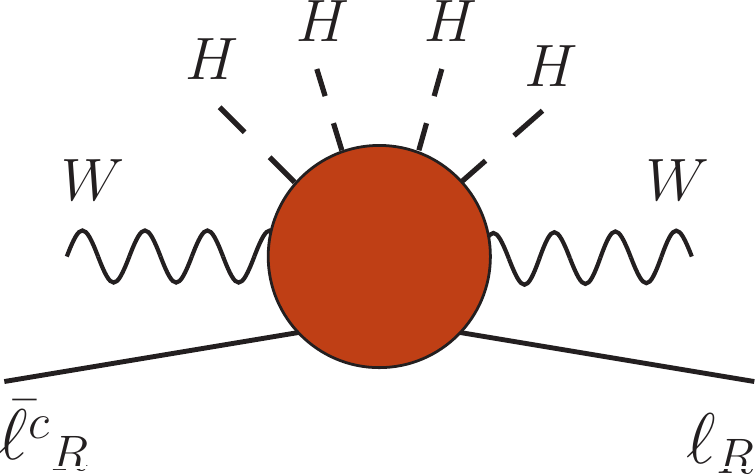} \hspace{2cm} 
\includegraphics[width=.45 \textwidth]{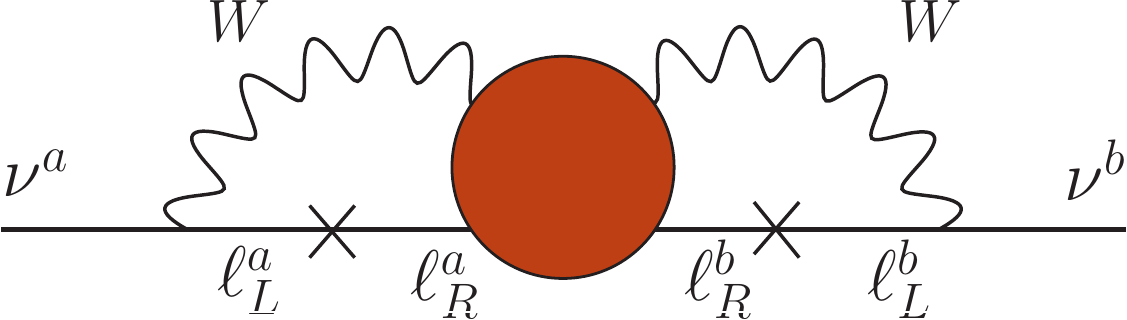}}
\caption{\small Left: Operator $\mathcal{O}_9$ from Eq.~(\ref{D9}). Right: 2-loop neutrino mass from $\mathcal{O}_9$. The crosses ($\times$) correspond to SM charged-lepton mass insertions.}
\label{Fig1}
\end{figure}

From an Effective Field Theory (EFT) perspective, Majorana neutrino mass generation is described via non-renormalizable (dimension $D \geq 5$) operators of SM fields that break lepton number by two units ($\Delta L = 2$). As outlined in the introduction, a general classification of such $SU(3)_C \times SU(2)_L \times U(1)_Y$ gauge invariant operators\footnote{We assume that no gauge symmetries beyond of those of the SM (and under which the SM fields are charged) are present.}, built from SM quarks, leptons and the SM Higgs doublet, is provided in~\cite{Babu:2001ex,deGouvea:2007qla} up to dimension $D = 11$. However, operators involving SM gauge bosons are not contained in this classification. Here we focus on one such $\Delta L = 2 $ operator involving SM gauge bosons (see Figure \ref{Fig1} (left)), given by 
%
\be
\label{D9}
\mathcal{O}_9 \equiv
\,\overline{\ell^c}_{R_a} \ell_{R_b} \left[\left(D_{\mu} H\right)^T i\sigma_2 H \right]^2\,,
\ee
with $a,b$ being flavor indices on the right-handed charged leptons $\ell_R$, $H$ being the SM Higgs doublet and $\sigma_{\alpha}$ (${\alpha}=1,2,3$) being the Pauli matrices.
This $D = 9$ operator is very particular, namely it is the unique lowest order $\Delta L = 2$ effective SM operator under the assumption that the new physics breaking lepton number couples only to SM leptons of right-handed chirality (and not to those of left-handed chirality)~\cite{delAguila:2012nu}.
Besides its uniqueness, $\mathcal{O}_9$ has several features which make it rather appealing for neutrino mass generation: 
{\it(i)} It generates neutrino masses at 2-loop order (see Figure~\ref{Fig1} (right)), yielding an elegant argument for the smallness of neutrino masses. 
{\it(ii)} In the flavor basis, the entries of the neutrino mass matrix $m^{\nu}_{ab}$ are proportional to the SM charged lepton masses: $m^{\nu}_{ab} \propto m^{\ell}_{a} m^{\ell}_{b}$, which results in specific, testable correlations among neutrino mixing parameters~\cite{Gustafsson:2014vpa}. 
{\it(iii)} It leads to a short-range contribution to neutrinoless double $\beta$-decay processes which dominates over the one from light Majorana neutrino exchange~\cite{Gustafsson:2014vpa}, to possibly be within reach of upcoming experiments.

\begin{figure}
\center{
\includegraphics[width=.3 \columnwidth]{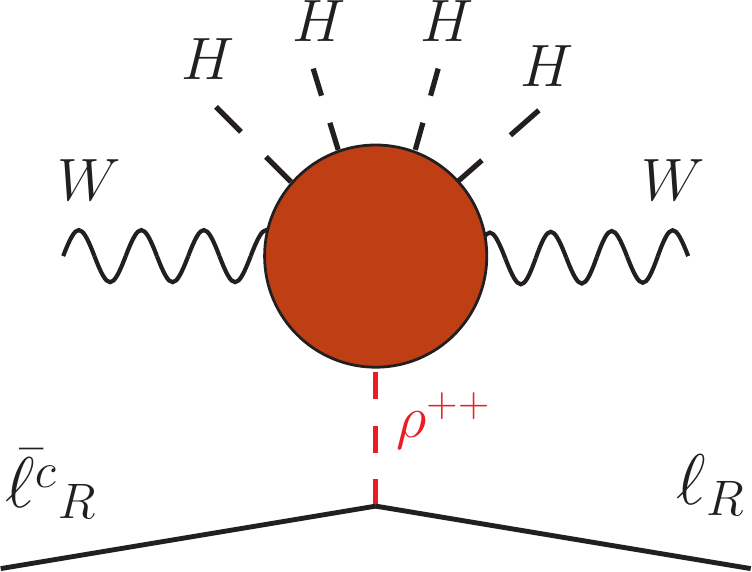}
\hspace{1.3cm} 
\includegraphics[width=.3 \columnwidth]{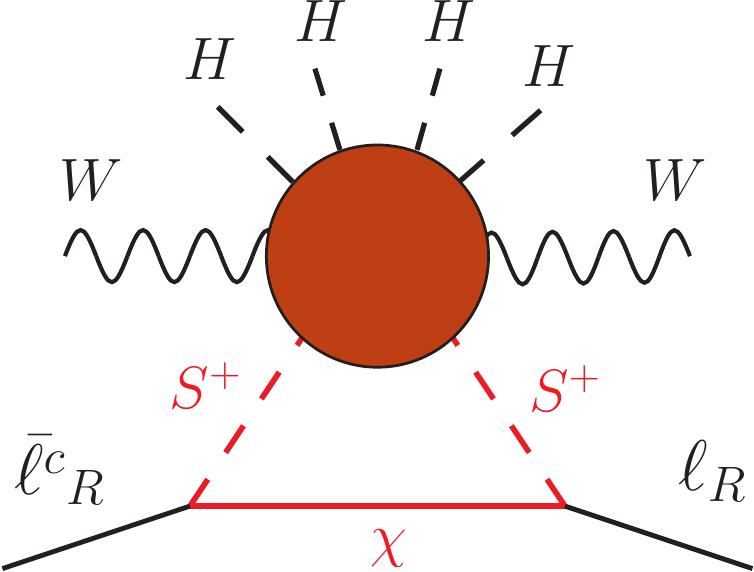}

\vspace{5mm} 

\includegraphics[width=.3 \columnwidth]{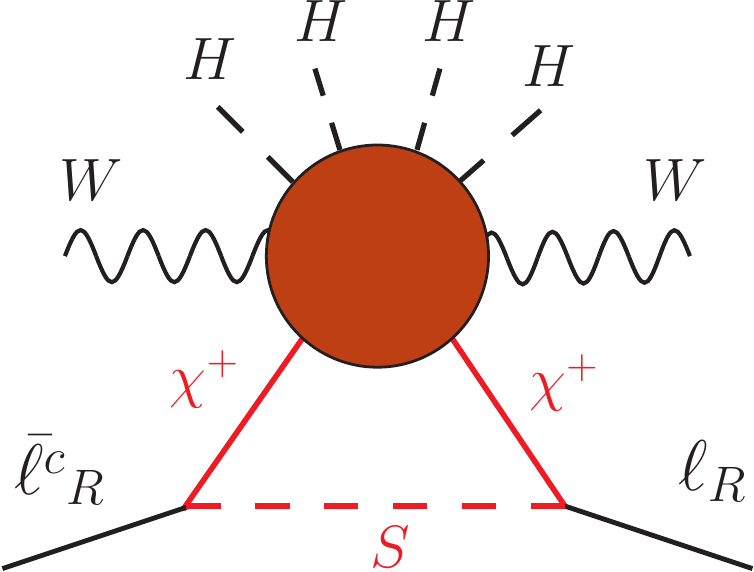}
\hspace{1.3cm}
\includegraphics[width=.3 \columnwidth]{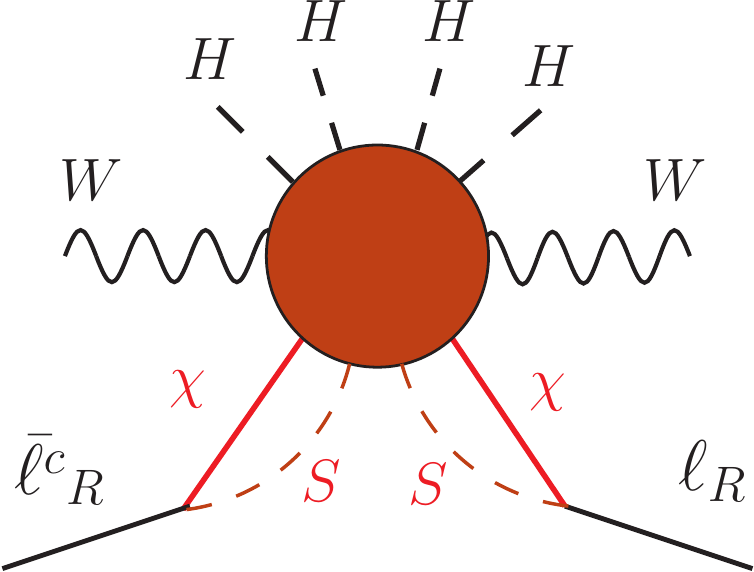}
}
\caption{\small 
Top-left: Opening-up operator $\mathcal{O}_9$ in Eq.~\eqref{D9} via a $\rho^{++}$ state,
giving rise to the effective BSM operator $\mathcal{O}^1_{\scriptscriptstyle{\text{BSM}}}$ of Eq.~\eqref{D7}. 
Top-right, Bottom-left, Bottom-right: Opening-up $\mathcal{O}_9$ via $S$ and $\chi$ states, respectively giving rise to the effective BSM operators $\mathcal{O}^\text{2a}_{\scriptscriptstyle{\text{BSM}}}$, $\mathcal{O}^\text{2b}_{\scriptscriptstyle{\text{BSM}}}$ and 
$\mathcal{O}^\text{2c}_{\scriptscriptstyle{\text{BSM}}}$
in Eq.~(\ref{O2a}-\ref{O2c}). BSM states are depicted in red.
}
\label{Fig2}
\end{figure}

\medskip
By inspecting Figure \ref{Fig1} (left) and Eq.~\eqref{D9}, there are only two possible classes of models that generate the effective operator $\mathcal{O}_9$ from a renormalizable theory:

\vspace{3mm}
\noindent $\bullet$ \textbf{Class 1:} 
The only possible renormalizable interaction involving the charged-lepton bilinear $\overline{\ell^c}_{R} \ell_{R}$ involves an $SU(2)_L$ singlet, doubly charged ($Y=2$) scalar $\rho^{++}$~\cite{Gustafsson:2014vpa,King:2014uha,Geib:2015tvt}. The corresponding operator is
\be
\label{D9UV_1}
C_{ab}\,\overline{\ell^c}_{R_a} \ell_{R_b} \,\rho + \mathrm{h.c.}
\ee
where $C_{ab}$ is a (complex) Yukawa matrix in flavor space. 
Trading $\overline{\ell^c}_{R} \ell_{R}$ in $\mathcal{O}_9$ for $\rho^c$ yields the dimension $D = 7$ operator~\cite{King:2014uha} (see Figure~\ref{Fig2} (top-left))
%
\be
\label{D7}
\mathcal{O}^1_{\scriptscriptstyle{\text{BSM}}} \equiv \,\rho^c \left[\left(D_{\mu} H\right)^T i\sigma_2 H \right]^2\,,
\ee
which together with Eq.~(\ref{D9UV_1}) breaks lepton number by two units and yields neutrino masses. The neutrino mass matrix reads
\be
\label{NuMassClass1}
m^{\nu}_{ab} = \frac{1}{(16\pi^2)^{L+2}}\, C_{ab}   \frac{m^{\ell}_a\, m^{\ell}_b}{\Lambda} \, , 
\ee
where $\Lambda$ is an effective energy scale (dependent also on $m_{\rho}$) and $L$ is the loop order of the ultraviolet (UV) completion of the operator $\mathcal{O}^1_{\scriptscriptstyle{\text{BSM}}}$.  
Clearly, the operator needs to be completed as $\rho$ cannot couple directly to $W$ bosons since it is not charged under $SU(2)_L$. Besides, the states (charged under $SU(2)_L$) needed to mediate the interaction between $\rho$ and the $W$ bosons in Eq.~(\ref{D7}) have to lie beyond the SM. The different models within this class then differ in the specific nature and properties of these BSM states.
In scenarios without a DM particle candidate, the interactions between $\rho$ and the $W$ bosons can occur at tree-level~\cite{Chen:2006vn,delAguila:2011gr}, yielding $L=0$ in Eq.~(\ref{NuMassClass1}). In contrast, for scenarios where the UV completion to $\mathcal{O}^1_{\scriptscriptstyle{\text{BSM}}}$ involves DM particles, the operator $\mathcal{O}_9$ is generated at 1-loop~\cite{Gustafsson:2012vj,Jin:2015cla,Alcaide:2017xoe} (or higher). This can be understood since a DM stabilizing symmetry, e.g. $\mathbb{Z}_2$, constrains the allowed interactions to achieve a UV completion, requiring in particular an even number of states charged under the $\mathbb{Z}_2$ symmetry. This then yields $L \geq 1$ in Eq.~(\ref{NuMassClass1}), with DM mediating the interactions between $\rho$ and the $W$ bosons at loop level. As a result, neutrino masses are generated in these scenarios at 3-loops (or beyond).

\vspace{3mm}
\noindent $\bullet$ \textbf{Class 2:} 
It is also possible to ``open-up'' the bilinear $\overline{\ell^c}_{R} \ell_{R}$  within $\mathcal{O}_9$. We introduce a new scalar $S$ and spin 1/2 fermion $\chi$, with the bilinear $\chi\, S$ being singlet under $SU(2)_L$ and with hypercharge $Y=1$. The SM right-handed charged lepton $\ell_{R}$ can now be coupled to this bilinear via
\be
\label{D9UV_2}
g_a\, \ell_{R_a} ( \chi \, S) + \mathrm{h.c.}
\ee
with $g_a$ being a Yukawa coupling. Regarding the hypercharge assignments for $S$ and $\chi$ within the $\chi \, S$ bilinear, one possibility is for either $\chi$ or $S$ to carry hypercharge $Y = 1$ (with the other field having $Y = 0$). The other possibility is for both $\chi$ and $S$ to have fractional hypercharge (e.g.~$Y_S = Y_{\chi} = 1/2$). In the former case, this allows to trade the bilinear $\overline{\ell^c}_{R} \ell_{R}$ in $\mathcal{O}_9$ for either $S S$ (if $S$ has hypercharge $Y =1$) or $\bar{\chi}^c \chi$ (if $\chi$ has hypercharge $Y = 1$), resulting respectively in the effective dimension 8 and 9 operators
\bea
\label{O2a}
\mathcal{O}^\text{2a}_{\scriptscriptstyle{\text{BSM}}} \equiv \,S S \left[\left(D_{\mu} H\right)^T i\sigma_2 H \right]^2\\
\label{O2b}
\mathcal{O}^\text{2b}_{\scriptscriptstyle{\text{BSM}}} \equiv \,\bar{\chi}^c \chi \left[\left(D_{\mu} H\right)^T i\sigma_2 H \right]^2
\eea
as shown in Figure~\ref{Fig2} (top-right, bottom-left). In the case where $S$ carries hypercharge $Y = 1$, in order for lepton number to be broken and neutrino masses to be generated --- the presence of (\ref{D9UV_2}) and (\ref{O2a}) is not enough (lepton number conserving charge assignments for $\chi$ and $S$ exist if only these two terms are present) --- a Majorana mass term for the fermion $m_\chi \bar{\chi}^c \chi$ must also be present. Alternatively, when $\chi$ carries hypercharge $Y = 1$ the combined presence of (\ref{D9UV_2}), (\ref{O2b}) and a mass term $m_s^2 S^2$ for the scalar is needed to generate neutrino masses. 
In addition, the bilinears $\bar{\chi}^c \chi$, $S S$, $\bar{\chi}^c\, S$  and $\chi\, S$ need all to be $SU(2)_L$ singlets for Eqs.~\eqref{D9UV_2},~\eqref{O2a} and~\eqref{O2b}~to be $SU(2)_L$ gauge invariant. Both $\chi$ and $S$ then need to transform under the same real representation of $SU(2)_L$. Thus, $\chi$ and $S$ are both $SU(2)_L$ $n$-tuplets with $n$ odd, e.g.~both singlets or both 
triplets\footnote{Higher $SU(2)_L$ representations~\cite{Chao:2018xwz} such as quintuplets or septuplets are also possible, yet viable completion of the operators $\mathcal{O}^\text{2a}_{\scriptscriptstyle{\text{BSM}}}$, $\mathcal{O}^\text{2b}_{\scriptscriptstyle{\text{BSM}}}$ or $\mathcal{O}^\text{2c}_{\scriptscriptstyle{\text{BSM}}}$ may become phenomenologically more difficult.}. 

When $S$ and $\chi$ have fractional hypercharges, the bilinear  $\overline{\ell^c}_{R} \ell_{R}$ in $\mathcal{O}_9$ can only be traded by the $Y = 2$ term $(\bar{\chi}^c \, S)(\chi \, S)$, yielding the dimension 11 operator (see Figure~\ref{Fig2} (bottom-right))
\bea
\label{O2c}
\mathcal{O}^\text{2c}_{\scriptscriptstyle{\text{BSM}}} \equiv \, (\bar{\chi}^c S) (\chi S) \left[\left(D_{\mu} H\right)^T i\sigma_2 H \right]^2
\eea
which together with (\ref{D9UV_2}) breaks lepton number by two units.
Now only the bilinears $\bar{\chi}^c\, S$  and $\chi\, S$ need to be $SU(2)_L$ singlets for Eqs.~\eqref{D9UV_2} and~\eqref{O2c} to be $SU(2)_L$ gauge invariant, which allows $\chi$ and $S$ to transform under complex representations~\cite{Chao:2018xwz} (in addition to the real representations discussed above) of $SU(2)_L$ as long as $\chi$ and $S$ transform in a conjugate representation of each other.

\medskip
The neutrino mass matrix stemming from operators $\mathcal{O}^\text{2a}_{\scriptscriptstyle{\text{BSM}}}$ and $\mathcal{O}^\text{2b}_{\scriptscriptstyle{\text{BSM}}}$ reads
\be
\label{NuMassClass2}
m^{\nu}_{ab} = \frac{1}{(16\pi^2)^{L+3}}\, g_{a} g_{b}   \frac{m^{\ell}_a\, m^{\ell}_b}{\Lambda'} \, , 
\ee
with $\Lambda'$ an effective scale and $L$ the loop order at which the operator $\mathcal{O}^\text{2a}_{\scriptscriptstyle{\text{BSM}}}$ or $\mathcal{O}^\text{2b}_{\scriptscriptstyle{\text{BSM}}}$ is UV completed by a renormalizable theory of neutrino masses, which requires new fields in addition to $\chi$ and $S$. 
As shown in Figure~\ref{Fig2} (top-right, bottom-left) neutrino masses will in this case be generated at three loops or higher. An exception occurs for $\mathcal{O}^\text{2b}_{\scriptscriptstyle{\text{BSM}}}$ when the neutral component of the field $S$ develops a vev, resulting in $\ell_{R}$ - $\chi$ mixing and neutrino mass generation at two loops or higher~\cite{delAguila:2012nu}.

In contrast, for operator $\mathcal{O}^\text{2c}_{\scriptscriptstyle{\text{BSM}}}$ neutrino masses are in general generated at a minimum of four loops, as apparent from Figure~\ref{Fig2} (bottom-right). The neutrino mass matrix then reads
\be
\label{NuMassClass2_bis}
m^{\nu}_{ab} = \frac{1}{(16\pi^2)^{L+4}}\, g_{a} g_{b}   \frac{m^{\ell}_a\, m^{\ell}_b}{\Lambda'} \, .
\ee
We note that, similarly to the $\mathcal{O}^\text{2b}_{\scriptscriptstyle{\text{BSM}}}$ operator, if the field $S$ has a neutral component and it develops a vev, this can result in neutrino masses being generated at two loops or higher ($L+2$). 
Apart from this exception, renormalizable models associated to $\mathcal{O}^\text{2c}_{\scriptscriptstyle{\text{BSM}}}$ 
will in general yield too small neutrino masses (suppressed by four or more loops), and we do not consider them further in this work.

\medskip
We also stress that for the above classes of models to give $\mathcal{O}_9$ as the leading lepton number violating operator, no lower order $\Delta L = 2$ operators must be generated by the new fields $\chi$ and $S$. 
which may require further ingredients depending on the specific quantum numbers of $\chi$ and $S$. An appealing possibility is to consider both fields to be odd under a $\mathbb{Z}_2$ symmetry. This automatically forbids interactions 
like e.g. $H\ell_L \chi$ (for an $SU(2)_L$ singlet $\chi$, which would lead to the generation of the $\Delta L = 2$ Weinberg $D = 5$ operator~\cite{Weinberg:1979sa}) or 
$H\chi\ell_R$ (for an $SU(2)_L$ doublet $\chi$). In addition, this $\mathbb{Z}_2$ symmetry would stabilize the lightest $\mathbb{Z}_2$-odd particle to provide possible DM particle candidates in these scenarios.

\medskip
This categorization that UV completions of the lepton number violating operator $\mathcal{O}_9$ from Eq.~\eqref{D9} fall into one out of two classes, defined respectively by containing the interaction \eqref{D9UV_1} or \eqref{D9UV_2}, is our first main result. 
In the next section we investigate concrete scenarios for neutrino mass generation belonging to {\sl Class 1} and {\sl Class 2}. We discuss the distinct phenomenological features of each class of models, as well as of each individual model within its class.  

\section{UV completions of the $\mathcal{O}_9$ operator and their phenomenology} \label{sec:UV_complete}

The phenomenology of renormalizable BSM completions of $\mathcal{O}_9$ is generically very rich, as a consequence of the required range of masses and couplings of the BSM fields to reproduce the observed pattern of neutrino masses and mixings, and includes potentially measurable lepton flavour violating processes, contributions to electroweak precision observables, signatures at the Large Hadron Collider (LHC) and future colliders, direct and indirect DM signals, as well as short-range contributions to neutrinoless double $\beta$-decay of atomic nuclei. We now discuss general phenomenological aspects as well as specific BSM scenarios for each class, and leave a detailed analysis of the interplay between neutrino mixing and neutrinoless double $\beta$-decay in {\sl Class 1} and {\sl Class 2} to section~\ref{sec:Constraints}. 

\subsection{Class 1}\label{sec:class.1}

The presence of the doubly charged scalar $\rho^{++}$ defines the general phenomenology of this class of (radiative) neutrino mass models. We note that when the interactions between $\rho$ and the $W$ bosons in Eq.~\eqref{D7} occur at tree-level, this is generally through the mixing of $\rho$ with the doubly charged component $\Delta^{++}$ of a scalar $SU(2)_L$ triplet $\Delta$ with $Y=1$ 
(see e.g.~\cite{Chen:2006vn,delAguila:2011gr}), which affects the properties of $\rho$. In this work we focus instead on the case where the properties of $\rho$ are not altered through its mixing with other states, yielding a loop-induced interaction between $\rho$ and the $W$ bosons. 

\begin{figure}
\center{
\includegraphics[width=.36 \columnwidth]{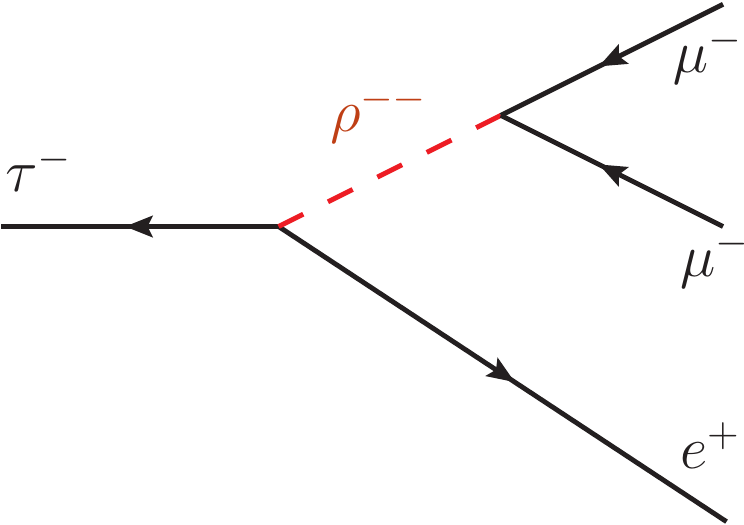} \hspace{1cm}
\includegraphics[width=.4 \columnwidth]{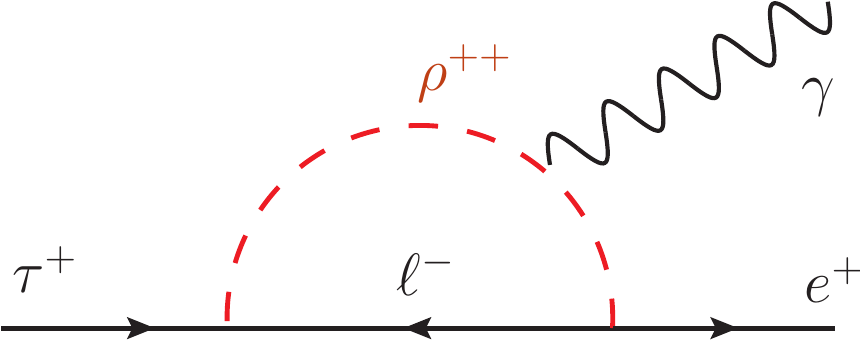}
}
\caption{\small Feynman diagrams in {\sl Class 1} for lepton flavour violating processes mediated by $\rho^{\pm\pm}$ at tree-level (Left: $\tau^{-}\rightarrow e^{+}\, \mu^{-}\, \mu^{-}$) and loop-level (Right: $\mu^+\rightarrow e^+ \gamma$).  
}
\label{fig:LFV}
\end{figure}

\medskip
The presence of the $\rho^{++}$ state has a wide impact on phenomenology.  
First, $\rho$ mediates contributions to various lepton flavour violating processes at tree and loop-level (see e.g.\ Figure~\ref{fig:LFV}).
In all cases, these contributions scale as $(C_{ab}/m_\rho)^2$. For a neutrino mass matrix of the form $m^{\nu}_{ab} \propto C_{ab} \,m^{\ell}_a \,m^{\ell}_b$ (as in Eq.~(\ref{NuMassClass1})) the matrix entries $m^\nu_{ee}$ and $m^\nu_{e\mu}$ are in general unavoidably tiny\footnote{The $m_{e}^2/v^2$ and $m_{e} m_{\mu}/v^2$ suppression cannot be compensated by altering any coupling constants or particle masses if they are to be kept within the perturbative and allowed regimes when $L> 0$.}. 
Compatibility with the observed pattern of neutrino mixing then requires all other entries in the neutrino mass matrix to be $m^{\nu}_{ab} \sim 0.01$ eV (see e.g. Figure~4 from~\cite{Gustafsson:2014vpa}).
This induces a generic Yukawa hierarchy (see also the discussion in section~\ref{sec:Constraints}) 
$C_{e\tau} \gg C_{\mu\mu} \gg C_{\mu\tau} \gg C_{\tau\tau}$~\cite{Gustafsson:2014vpa,King:2014uha}. As a result, 
the current most important constraints on the doubly charged scalar $\rho$ from lepton flavour violation 
are (see e.g.\ \cite{King:2014uha,Herrero-Garcia:2014hfa,Crivellin:2018ahj,Cepedello:2020lul})

\vspace{1mm}

\hspace{-11mm}
\begin{tabular}{llrcl}
$\mu^+\rightarrow e^+ \gamma$ & \hspace{3mm} $\mathrm{BR}<4.2 \times 10^{-13}$~\cite{TheMEG:2016wtm} 
&$\longrightarrow$& $\big|\sum_\ell C_{\ell\mu}\,C_{\ell e}^* \big|$ & \hspace{-4mm} $ < 2.4 \times 10^{-4} \;(m_{\rho}/\mathrm{TeV})^2$ \\
$\tau^{-}\rightarrow e^{+}\, \mu^{-}\, \mu^{-}$&\hspace{3mm} $\mathrm{BR}<9.8 \times 10^{-9}$~\cite{Amhis:2016xyh,Lees:2010ez,Hayasaka:2010np}
&$\longrightarrow$& $| C_{e\tau}\,C_{\mu\mu} | $& \hspace{-4mm} $ < 4.1 \times 10^{-3} \;(m_{\rho}/\mathrm{TeV})^2$ \\
$\tau^{-}\rightarrow e^{+}\, e^{-}\, \mu^{-}$ &   \hspace{3mm} $\mathrm{BR}<1.6 \times 10^{-8}$~\cite{Amhis:2016xyh,Hayasaka:2010np,Lees:2010ez} 
&$\longrightarrow$& $| C_{e\tau}\,C_{e\mu} | $ & \hspace{-4mm} $ < 4.6 \times 10^{-3} \;(m_{\rho}/\mathrm{TeV})^2$  \\
$\mu^-\rightarrow e^+ \,e^{-}\, e^{-} $	& \hspace{3mm} $\mathrm{BR}<1.0 \times 10^{-12}$~\cite{Bellgardt:1987du}  
&$\longrightarrow$ & $ | C_{e\mu}\,C_{ee} | $	& \hspace{-4mm} $ < 2.3 \times 10^{-5} \,(m_{\rho}/\mathrm{TeV})^2$  
\end{tabular} 
\label{LFV_Class1}

\vspace{1mm}
\noindent For specific models within {\sl Class 1}, some of the above lepton flavour violation rates can 
become large, which generally leads to stringent constraints within these models~\cite{Cepedello:2020lul}.

\begin{figure}
\center{\includegraphics[width=.73 \columnwidth]{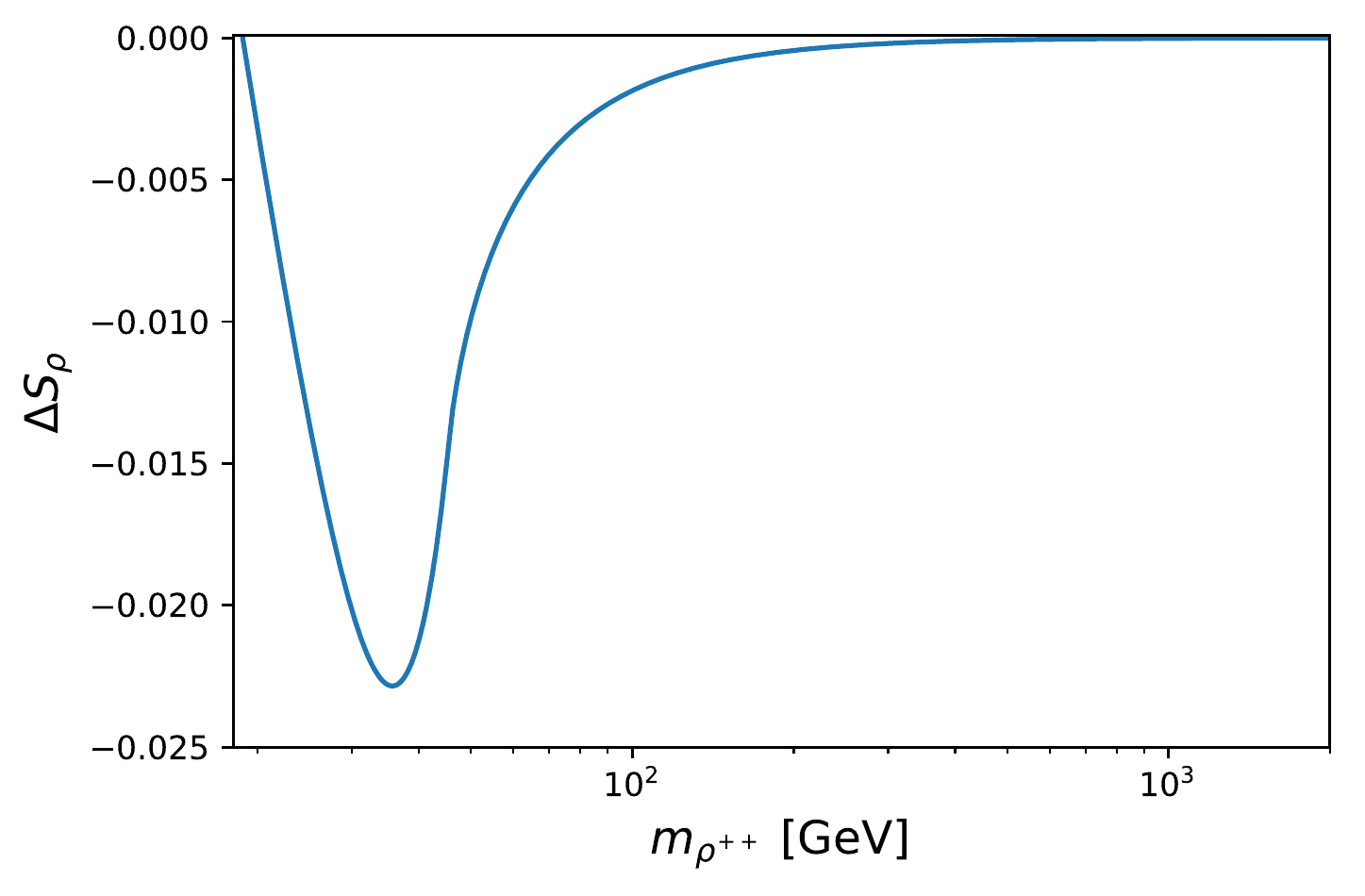}}
\vspace{-3mm}
\caption{
\small Contribution to the oblique parameter $S$ from the $SU(2)_L$ singlet $\rho^{++}$ as a function of its mass $m_{\rho}$. Current experimental result is $S=0.02\pm0.10$ \cite{Tanabashi:2018oca}, with $S=0$ from the SM.}  
\label{fig:STU_lolypop}
\end{figure}

\medskip
The doubly charged scalar $\rho^{++}$ contributes to electroweak precision observables through contributions to the oblique parameters $S$, $T$, $U$ \cite{Peskin:1991sw}. Being an $SU(2)_L$ singlet, its contribution to the parameter $T$ vanishes, but not those to the $S$ and $U$ parameters. In Figure~\ref{fig:STU_lolypop} we show the dependence on the $S$ parameter with the $\rho$ mass, finding that its contribution to the $S$ parameter is well within present experimental constraints. Likewise, $\rho$'s contribution $U_\rho = -S_\rho$ \cite{Grimus:2008nb} remains within the experimental bound $\Delta U \in \{-0.09,0.09\}$ \cite{Tanabashi:2018oca}.
We note that additional contributions to the oblique parameters will come from the fields that UV complete the $\mathcal{O}^1_{\scriptscriptstyle{\text{BSM}}}$ operator; these will be discussed below in the context of specific models.

\medskip
Finally, $\rho$ can be searched for at the LHC through Drell-Yan pair production~\cite{delAguila:2013yaa,delAguila:2013mia,Geib:2015tvt} $p p \to \gamma^*,Z^{*} \to \rho^{++}\rho^{--}$, with subsequent decays of the doubly charged scalars into di-leptons. Decays $\rho^{\pm\pm} \to W^{\pm}W^{\pm}$ are also possible~\cite{King:2014uha,delAguila:2013yaa,delAguila:2013mia,Kanemura:2013vxa,Kanemura:2014goa}, as well as cascade decays into BSM states (see e.g.~\cite{Alcaide:2017dcx}). Regarding the leptonic decays of $\rho^{++}$, $C_{e\tau}$ is generically the largest entry in the Yukawa matrix $C_{ab}$, and thus the dominant di-leptonic decay mode is $\rho^{\pm\pm} \to e^{\pm}\tau^{\pm}$ in this class of scenarios. This has two important consequences: {\it(i)} LHC searches for doubly charged scalars are currently tailored to the decays $\rho^{\pm\pm} \to e^{\pm}e^{\pm},\,e^{\pm}\mu^{\pm},\,\mu^{\pm}\mu^{\pm}$~\cite{ATLAS:2012hi,ATLAS:2016pbt}. The current limit $m_{\rho} > 420$ GeV for an $SU(2)_L$-singlet scalar $\rho^{\pm\pm}$~\cite{ATLAS:2016pbt} gets significantly weakened if the decay $\rho^{\pm\pm} \to e^{\pm}\tau^{\pm}$ is dominant, due to the much lower efficiencies for $\tau$-leptons in the final state~\cite{delAguila:2013mia}. Masses down to $m_{\rho} \sim 150$ GeV could in principle be allowed. {\it(ii)} The dominant decay $\rho^{\pm\pm} \to e^{\pm}\tau^{\pm}$ allows to distinguish this scenario at the LHC and future colliders from others with doubly charged scalars, e.g.\ the Zee-Babu model~\cite{Babu:1988ki, Nebot:2007bc} where the dominant decay is $\rho^{\pm\pm} \to \mu^{\pm}\mu^{\pm}$ or Type-II see-saw scenarios (see the discussion in~\cite{Schmidt:2014zoa}).

The collider bounds on $m_\rho$ imply that the effective scale $\Lambda$ in Eq.~\eqref{NuMassClass1} becomes of similar (or larger) magnitude than such bounds. Sizable Yukawa coupling will then be required to achieve correct neutrino mass matrix entries, e.g.\ for $m_{e\tau} \simeq 0.01$~eV and $\Lambda \simeq 100$~GeV, Eq.~\eqref{NuMassClass1} implies $C_{e\tau} \gtrsim (16\pi^2)^{(L-1)}$.\footnote{Attention might also be required on that perturbativity is ensured in these scenarios.}
As a result of this combination of unavoidably sizable couplings and not too large masses of BSM states, a rich phenomenology is expected in any particular model realization within this class, with strong constraints on the parameter space arising from the interplay between low and high-energy observables \cite{Gustafsson:2012vj,Alcaide:2017xoe,King:2014uha,Geib:2015tvt,Cepedello:2020lul}.
In the following we discuss in detail two renormalizable models belonging to {\sl Class 1} which yield neutrino masses at 3-loops and include a DM particle.

\subsubsection{Model 1: Inert Doublet Dark Matter (The {\it Cocktail} Model)}
\label{Sec_cocktail}

This model, originally proposed in~\cite{Gustafsson:2012vj}, contains a scalar $SU(2)_{L}$ doublet $\Phi$ and a charged scalar $SU(2)_{L}$ singlet $S^{\pm}$ to complete the $\mathcal{O}^1_{\scriptscriptstyle{\text{BSM}}}$ operator of Eq.~\eqref{D7}. Both fields $\Phi$ and $S^{+}$ are assigned to be odd under an unbroken $\mathbb{Z}_2$ symmetry, guaranteeing that no SM lepton number violating operator with $D < 9$ exists. The SM fields and $\rho^{++}$ are even under the $\mathbb{Z}_2$ symmetry. The Lagrangian for such a setup reads\footnote{The additional interactions from the scalar potential $V(H,\,\Phi, \,S,\, \rho)$ are left out for brevity, as they play no significant role in our discussion.}
\bea
\label{eq:LCM}
\mathcal{L} &\supset& 
(D_\mu \Phi)^\dagger D^\mu \Phi + (D_\mu S)^\dagger D^\mu S  + (D_\mu \rho)^\dagger D^\mu \rho 
-\frac{\lambda_5}{2} \left(H^{\dagger} \Phi \right)^2 \\&&
- \kappa_1 \, \Phi^{T} i \sigma_2 H\, S^{-}  
- \kappa_2 \, \rho^{++} S^{-} S^{-} 
- \xi \, \Phi^{T} i \sigma_2 H\, S^{+} \, \rho^{--} 
- C_{ab} \, \overline{\ell^c}_{R_a} \ell_{R_b} \, \rho^{++} +  \mathrm{h.c.}
\nonumber
\eea
where $a,b$ are the flavour indices on the right-handed charged leptons $\ell_R$ and $C_{ab}$ is a (complex) Yukawa matrix. The SM Higgs field $H$ and $\Phi$ read 
\be
H = \frac{1}{\sqrt{2}}\left(\begin{array}{c}
0 \\
h
\end{array}
\right) 
+
\left(\begin{array}{c}
0 \\
v
\end{array}
 \right) \quad, \quad
\Phi =\frac{1}{\sqrt{2}} \left(
\begin{array}{c}
\Lambda^{+} \\
H_0 + i\, A_0
\end{array} \right),
\ee
where $v \simeq 174$ GeV 
is the Higgs vacuum expectation value. $H$ and $\Phi$ will not mix due to their different $\mathbb{Z}_2$ symmetry properties. The lightest $\mathbb{Z}_2$-odd state is stable, and if this is one of the two neutral states $A_0$, $H_0$ then the model has a viable DM candidate. In the following we will consider $H_0$ to be the DM particle.

\medskip
After electroweak symmetry breaking the charged states $\Lambda^+$ and $S^+$ will mix if $\kappa_1 \neq 0$ to give the mass eigenstates 
\bea
 H^+_1 &=&   S^+ \sin\beta \,   + \,  \Lambda^+ \cos\beta \nonumber \\
 H^+_2 &=&   S^+ \cos\beta \,  -  \,  \Lambda^+ \sin\beta \,.
\eea
with the mixing angle being $\beta$.
A convenient set of independent parameters for the model are the masses for the five states $\rho$, $H_0$, $A_0$, $H_{1,2}^\pm$, the mixing angle $\beta$, the couplings $\xi$, $\kappa_{2}$ and the Yukawa matrix $C_{ab}$. In addition, the remaining parameters of the scalar potential $V(H,\,\Phi, \,S,\, \rho)$ should be properly chosen to preserve vacuum stability (see e.g.~\cite{Babu:2002uu}) and leave the $\mathbb{Z}_2$ symmetry unbroken.

\begin{figure}
\center{\includegraphics[width=0.54 \columnwidth]{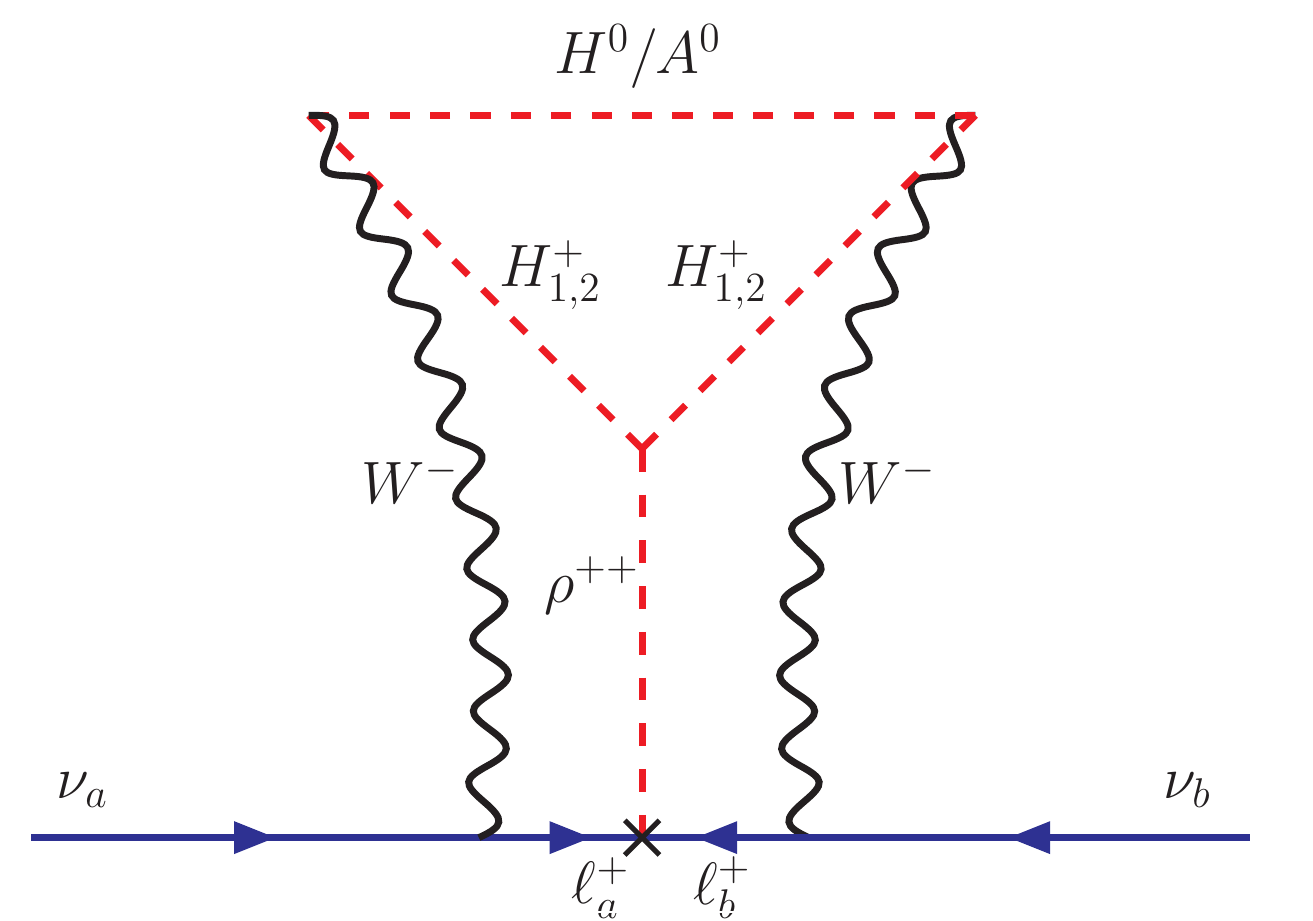}}
\caption{The ``cocktail diagram'' yielding neutrino masses in the {\sl Cocktail} model. BSM propagators are depicted in red.} 
\label{fig:Cocktail}
\end{figure}

The Lagrangian~\eqref{eq:LCM} breaks lepton number by two units if both $C_{ab}$ and $\lambda_5$ are non-zero together with either $\kappa_1,\,\kappa_2 \neq 0$, $\kappa_1,\,\xi \neq 0$ or $\kappa_2,\,\xi \neq 0$. In the first two cases the leading contribution to neutrino masses appears at 3-loop order -- via the so-called ``cocktail diagram'' shown in Figure~\ref{fig:Cocktail} (in contrast, in the latter case the leading contribution to neutrino masses appears at the 5-loop level! This would make the model not phenomenologically viable). The 3-loop neutrino mass matrix $m^{\nu}_{ab}$ has been calculated in \cite{Gustafsson:2012vj,Geng:2014gua} and is given by
\be
\label{eq:MajoranaMatrix}
m^{\nu}_{ab} = C_{ab}   \frac{m^{\ell}_a\, m^{\ell}_b \, s_{2\beta}}{(16\pi^2)^3 \,m_{\rho}} \, \frac{m_W^4\,\Delta m_+^2\,\Delta m_0^2}{v^8}\,
(\mathcal{A}_1 \, \mathcal{I}_1 + \mathcal{A}_2 \, \mathcal{I}_2), 
\ee 
where $m^{\ell}_a$, $m^{\ell}_b$ are the SM charged lepton masses (of flavour $a,\,b$), $s_{2\beta} \equiv \mathrm{sin} (2\beta)$, $m_W = 80.35$ GeV is the mass of the $W$-boson, $\Delta m_+^2 \equiv m^2_{H^{\pm}_2} - m^2_{H^{\pm}_1} \propto \kappa_1$ and $\Delta m_0^2 \equiv m^2_{A_0} - m^2_{H_0} \propto \lambda_5$. The coefficients $\mathcal{A}_{1,2}$ are approximately given by
\be
\mathcal{A}_{1} \simeq \frac{\kappa_2 \,s_{2\beta} + \xi\,v\,c_{2\beta}}{\sqrt{v\,m_{\rho}}}   \quad , \quad \mathcal{A}_{2} \simeq 10\, \xi,
\ee
and the integrals $\mathcal I_{1,2}$ are shown to be at most $\mathcal{O}(1)$ numbers when the masses of the new states are kept above current bounds from colliders and couplings are kept in a perturbative regime (see \cite{Gustafsson:2012vj,Geng:2014gua} for more details).

\smallskip
Besides the presence of the doubly charged scalar $\rho^{++}$, whose most relevant phenomenology has been discussed in section {\ref{sec:class.1}}, the states $H^+_{1,2}$, $A_0$, $H_0$ and the connection between neutrino masses and DM yield a wide range of phenomenological implications, which we summarize in the following:

\vspace{2mm}
\noindent \textsl{\textbf{Electroweak Precision Observables:}} 
Generating neutrino masses of the right size in the {\sl Cocktail} model requires large mass splittings among the $\mathbb{Z}_2$-odd states $H_{1,2}^{+}$, $A_0$, $H_0$, as $m^{\nu}_{ab}$ is proportional to $\Delta m_+^2$ and $\Delta m_0^2$ (see Eq.~\ref{eq:MajoranaMatrix}). These large splittings $\Delta m_+^2, \, \Delta m_0^2 \gtrsim v^2$ are strongly constrained by measurements of the electroweak oblique parameters, particularly the $T$-parameter (see e.g.~\cite{Belyaev:2016lok}). Thus, partial cancellations among different contributions to the oblique parameters are needed in the model. This yields specific mass correlations between the new states~\cite{Gustafsson:2012vj}, $m_{H^+_2} \gg m_{A_0} \gg m_{H^+_1}, m_{H_0}$ (with $m_{H^+_2}-m_{H^+_1} \gg v$ and $m_{A_0}-m_{H_0} \gtrsim v$).

\vspace{2mm}
\noindent \textsl{\textbf{Dark Matter:}} 
The DM phenomenology in the {\sl Cocktail} model is that of the well-studied Inert Doublet model~\cite{Ma:2006km,Barbieri:2006dq,LopezHonorez:2006gr} (see also~\cite{Gustafsson:2012aj,Arhrib:2013ela,Ilnicka:2015jba,Garcia-Cely:2015khw,Belyaev:2016lok,Eiteneuer:2017hoh} and references therein). Generating neutrino masses of the right size yields three possible scenarios for obtaining the correct 
DM relic abundance: {\it(a)} $m_{H_0} \simeq m_h/2$, the correct DM abundance obtained via annihilation into SM fermions through a resonant SM Higgs boson.~{\it(b)} $H_0 - H^{+}_1$ coannihilations for $m_{H_0}\sim 50 - 75$ GeV and $m_{H^{+}_1} - m_{H_0} \lesssim$ (few) GeV~\footnote{This region is constrained by LEP and LHC searches for charginos. A reinterpretation of these limits for the Inert Doublet Model has been performed in~\cite{Lundstrom:2008ai,Belanger:2015kga}, yielding $m_{H^+} \gtrsim 70$ GeV. In addition, LHC searches for an invisible width of the 125 GeV Higgs boson yield strong constraints on the $m_{H_0} < 62$~GeV region (see e.g.~\cite{Kalinowski:2018ylg}).}.~{\it (c)} $m_{H_0} \lesssim m_W$, where the closeness to the $W W$ threshold regulates the annihilation rate at freeze-out. All three scenarios remain viable given the current LHC and DM experimental constraints on the model~\cite{Ilnicka:2015jba,Belyaev:2016lok,Garcia-Cely:2015khw,Eiteneuer:2017hoh}.
DM direct detection experiments set a strong bound on the DM coupling to the SM Higgs $\lambda_{H_0}$, currently $\lambda_{H_0} \lesssim 0.03$ for $m_{H_0} < 100$ GeV from the latest XENON1T results~\cite{Aprile:2018dbl} (following from an extrapolation of the results in~\cite{Belyaev:2016lok}).
In addition, the model could produce a striking monochromatic gamma-ray line~\cite{Gustafsson:2007pc} detectable by the FERMI - Large Area Telescope (see~\cite{Garcia-Cely:2015khw,Eiteneuer:2017hoh} for a recent analysis of the Inert Doublet Model in light of indirect DM detection data).

\vspace{2mm}
\noindent \textsl{\textbf{LHC \& Future Collider Phenomenology:}} 
The LHC phenomenology of the $\mathbb{Z}_2$-odd states $H_{1,2}^{+}$, $A_0$, $H_0$ is similar to that of the Inert Doublet Model, and can lead to multilepton signatures~\cite{Gustafsson:2012aj} and jet(s) + $E_T^{\mathrm{miss}}$ signatures~\cite{Belyaev:2016lok,Dutta:2017lny,Belyaev:2018ext}.
Given the large mass splittings required in the model, the Drell-Yan production processes $p p \to A_0 H_0 \to H_0 H_0 Z$, $Z \to \ell\ell$ (mono-$Z$), 
$p p \to H^{\pm}_{1} H_0$ and $p p \to H^{\pm}_{1} H^{\mp}_{1}$ yield the most promising search avenues at the LHC and future collider facilities (we note that jet(s) + $E_T^{\mathrm{miss}}$ signatures strongly depend on the value of $\lambda_{H_0}$, currently very constrained by DM direct detection). Current constraints from LEP and the LHC are rather weak~\cite{Lundstrom:2008ai,Belanger:2015kga} 
(see also~\cite{Ilnicka:2015jba,Belyaev:2016lok}), e.g.\ the charged state $H^+_1$ is only constrained to lie above $\sim 70$ GeV. In addition, the presence of $H^{+}_{1}$ ($H^{+}_{2}$ is too heavy in general) provides a new possible decay channel $\rho^{\pm\pm} \to H^{\pm}_{1}H^{\pm}_{1}$, which would yield distinctive high-multiplicity lepton signatures via Drell-Yan pair production of $\rho$ (see e.g.~\cite{Alcaide:2017dcx}). Finally, the various charged BSM states in the model yield a contribution to $h\rightarrow\gamma\gamma$ (see e.g.~\cite{Arhrib:2012ia,Krawczyk:2013pea}), which can be important if any of the  BSM charged states has a mass around or below $100$~GeV.

\vspace{2mm}
\noindent \textsl{\textbf{Lepton Flavour Violation:}} 
Neutrino masses of the right size within the model require $C_{e\tau}/m_\rho \gtrsim 0.1/$TeV, $C_{\mu\mu}/m_\rho \gtrsim 10^{-2}/$TeV and $C_{\mu\tau}/m_\rho \gtrsim 10^{-3}/$TeV~\cite{Gustafsson:2012vj}. As a consequence, the lepton flavour violating processes $\mu \to e\, \gamma$ and $\tau^- \to e^+ \mu^{-} \mu^-$ strongly constrain the model~\cite{Cepedello:2020lul}. By inspecting the various present bounds for lepton flavour violation processes in Table~\eqref{LFV_Class1}, an order of magnitude improvement in current lepton flavour violation sensitivities would then lead to detectable signals or rule out the model.

\subsubsection{Model 2: Inert Triplet Dark Matter (The {\it Lollipop} model)}
\label{sec:lollipop}

This model has been proposed and analyzed in~\cite{Alcaide:2017xoe}. Besides the SM fields and the doubly charged scalar $\rho^{++}$, it includes a scalar $SU(2)_{L}$ triplet $\Delta$ with hypercharge $Y = 1$ and a real singlet scalar $\sigma$ in order to enable a UV completion of the $\mathcal{O}^1_{\scriptscriptstyle{\text{BSM}}}$ operator in Eq.~\eqref{D7}. As for the previous ({\sl Cocktail}) model, both fields $\Delta$ and $\sigma$ are odd under an unbroken $\mathbb{Z}_2$ symmetry, which guarantees that there is no SM lepton number violating operator with $D < 9$. The relevant part of the Lagrangian reads
\bea
\label{eq:LLM}
\mathcal{L} &\supset& 
\textnormal{Tr} [(D_\mu \Delta)^\dagger D^\mu \Delta] + \partial_\mu \sigma\, \partial^\mu \sigma + (D_\mu \rho)^\dagger D^\mu \rho 
- m^2_{\Delta} \textnormal{Tr}[\Delta^{\dagger} \Delta] - \frac{m^2_{\sigma}}{2} \sigma^2 \nonumber\\&& 
- \lambda_{H\sigma} \left|H\right|^2 \sigma^2
- \lambda_{H\Delta} \left|H\right|^2 \textnormal{Tr}[\Delta^{\dagger} \Delta]
- \tilde{\lambda}_{H\Delta} H^{\dagger} \Delta \Delta^{\dagger} H - \kappa_2 \textnormal{Tr}[\Delta \Delta]\rho^{--} \nonumber\\&&
- \lambda_6 \,\sigma H^{\dagger} \Delta \tilde{H} - C_{ab} \, \overline{\ell^c}_{R_a} \ell_{R_b} \, \rho^{++} +  \mathrm{h.c.}   \,,
\eea
where $H$ is the SM Higgs doublet and ${\tilde H} = i \sigma_2 H^*$ (the additional interactions from the scalar potential $V(\sigma,\rho,H,\Delta)$ are omitted as they will not play a role in the following discussion). The scalar triplet $\Delta$ is given by
\be
\label{scalar_triplet}
\Delta = \frac{1}{\sqrt{2}}
\left(
\begin{array}{c c}
\Delta^+      & \sqrt{2}\Delta^{++}\\
\Delta_0+iA_0 & -\Delta^+
\end{array}
\right).
\ee
After electroweak symmetry breaking the mass hierarchy among $A_0$, $\Delta^{+}$ and $\Delta^{++}$ is governed by the sign of $\tilde{\lambda}_{H\Delta}$ in Eq.~\eqref{eq:LLM}:
\be
m^2_{A_0} - m^2_{\Delta^+} = m^2_{\Delta^+} - m^2_{\Delta^{++}} = \tilde{\lambda}_{H\Delta} \, v^2/2
\ee
such that there is either $m_{A_0} > m_{\Delta^+} > m_{\Delta^{++}}$ or $m_{\Delta^{++}} > m_{\Delta^+} > m_{A_0}$. It can also be seen from Eq.~\eqref{eq:LLM} that lepton number violation in this scenario requires $C_{ab},\,\kappa_2,\,\lambda_6 \neq 0$. After electroweak symmetry breaking, $\lambda_6$ induces a mixing between $\Delta_0$, the neutral scalar part of the triplet field, and $\sigma$. This singlet-triplet mixing, $\mathrm{sin}\, \alpha$, yields two mass eigenstates $S_1$, $S_2$, with masses
\be
\label{eigenmasses_triplet}
m^2_{1,2} = \frac{1}{2} \left[m^2_{A_0} + \bar{m}^2_{\sigma} \pm \sqrt{(m^2_{A_0} - \bar{m}^2_{\sigma})^2 + 8 \lambda_6^2 v^4} \right]
\ee
with $\bar{m}^2_{\sigma} = m^2_{\sigma} + 2\,\lambda_{H\sigma} v^2$. From Eq.~\eqref{eigenmasses_triplet} we have $m_{S_1} > m_{A_0} > m_{S_2}$, with the splitting between $A_0$ and the mostly triplet-like mass eigenstate (from $S_1$, $S_2$) being controlled by $\lambda_6$. The state $S_2$ is the lightest $\mathbb{Z}_2$-odd state, and thus a viable DM candidate, provided that $m_{A_0}^2 - m^2_{S_2} > \tilde{\lambda}_{H\Delta} v^2$. 

\begin{figure}[t]
\begin{centering}
\includegraphics[width=0.42 \columnwidth]{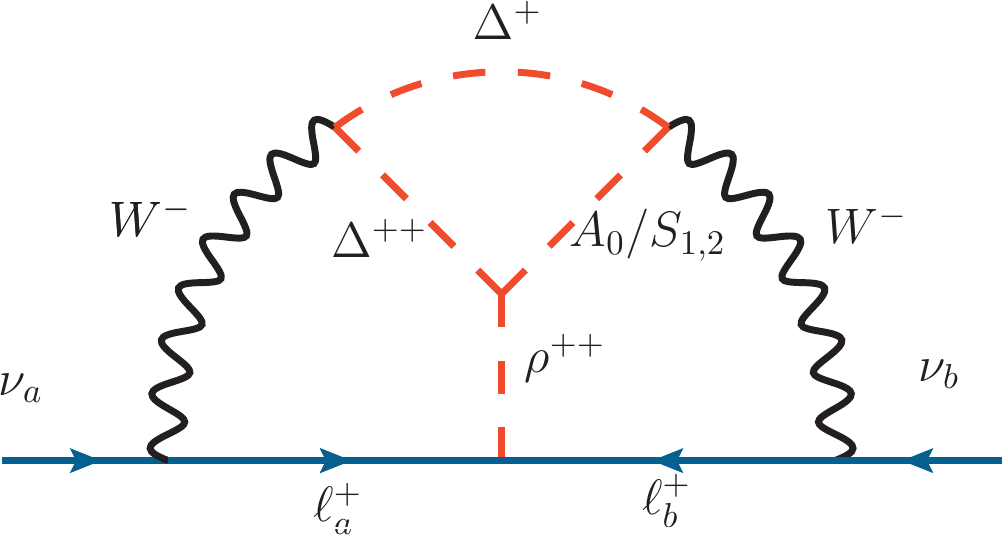}
\includegraphics[width=0.42 \columnwidth]{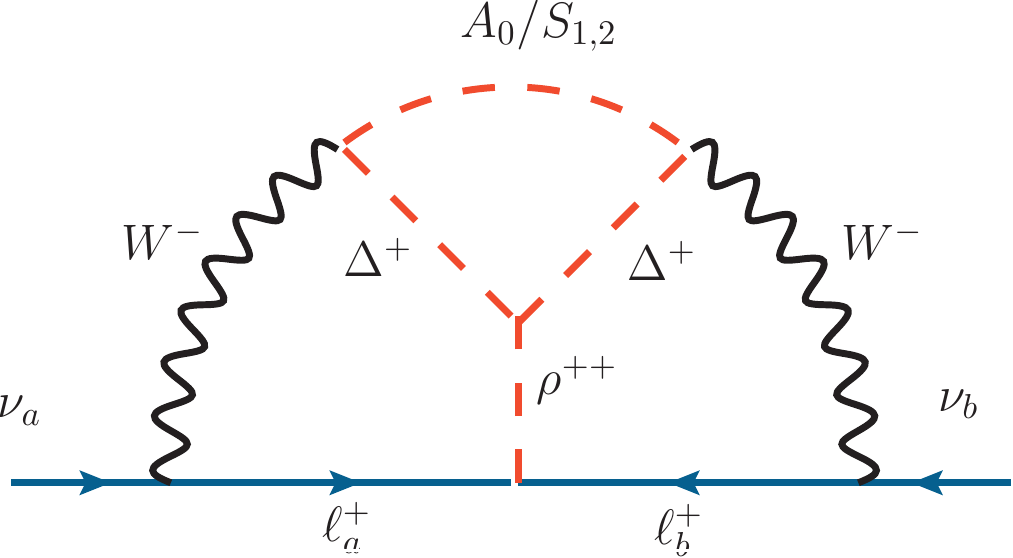}
\includegraphics[width=0.42 \columnwidth]{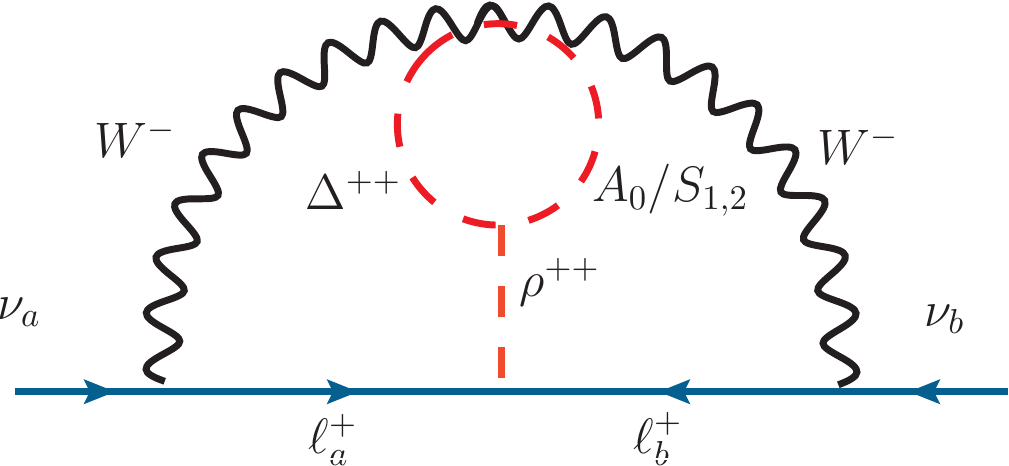}
\par\end{centering}
\caption{Example Feynman diagrams yielding neutrino masses in the Inert Triplet Dark Matter Model. BSM propagators are depicted in red.
\label{fig:lollypop}}
\end{figure}

Similarly to the previous scenario, the leading contributions to neutrino masses (shown in Figure~\ref{fig:lollypop}) appear first at 3-loop order. The neutrino mass matrix $m^\nu_{ab}$ has been calculated in \cite{Alcaide:2017xoe} and is given by
\bea
\label{eq:MajoranaMatrixLoll}
m^\nu_{ab}= C_{ab}\frac{m^{\ell}_a\, m^{\ell}_b}{(16\pi^2)^3 m^2_{\rho}}\,8\,\kappa_2 \lambda^2_6 \times \mathcal{I}_\nu
\eea
with $\mathcal{I}_\nu$ an integral that takes a value of $\mathcal{O}(1-10)$ provided the new states are close to the electroweak scale~\cite{Alcaide:2017xoe}.
Since the observed neutrino masses and mixing pattern require the entries in the neutrino mass matrix (except for $m^{\nu}_{ee}$ and $m^{\nu}_{e\mu}$) to lie in the range $m^{\nu}_{ab} \sim 0.01 - 0.03$ eV (see, e.g.\ \cite{Gustafsson:2014vpa}), generating neutrino masses of the right size from Eq.~\eqref{eq:MajoranaMatrixLoll} already requires $\lambda_6 \gtrsim 1$ and $\kappa_2 \gtrsim 1$ TeV. We also note that, if $m_{\rho}$ is to lie below the TeV scale, $\kappa_2$ cannot be pushed much above the TeV scale since it yields a 1-loop contribution to $m_{\rho}$ of the order $\delta m_{\rho}^2 \sim \kappa_2^2/(4\pi)^2$~\cite{Nebot:2007bc}.   

\vspace{2mm}

The phenomenology of this scenario is also very rich, bearing much resemblance to that of the previous model from section~\ref{Sec_cocktail}. Both are similar regarding lepton flavour violation, as the doubly charged scalar $\rho^{++}$ is the only BSM state interacting with the SM leptons\footnote{Yet, since the allowed values of the $\rho^{++}$ mass and its Yukawa couplings to SM leptons may differ from those of the {\sl Cocktail} model, the predicted strengths of lepton flavour violation processes may be different in the two models.}. However, the DM phenomenology in this scalar triplet DM scenario is notably different from that discussed in section~\ref{Sec_cocktail}, and the additional presence of the doubly charged scalar $\Delta^{++}$ in this model can have important implications for electroweak precision observables and LHC physics. A brief overview of the model's phenomenology is given below:

\vspace{1mm}

\noindent \textsl{\textbf{Electroweak Precision Observables:}} Besides the $\rho^{++}$ contribution to the oblique parameters, which has been discussed above, all the $\mathbb{Z}_2$-odd states $\Delta^{+}$, $\Delta^{++}$, $A_0$, $S_{1,2}$ yield contributions to the oblique parameter $T$ (as discussed in \cite{Alcaide:2017xoe}), as well as to $S$ and $U$. We give explicit expressions for the $S$, $T$ and $U$ oblique parameters in this model in Appendix~\ref{app:A}. 
\begin{figure}
\center{\includegraphics[angle=0, width=1.0
\columnwidth]{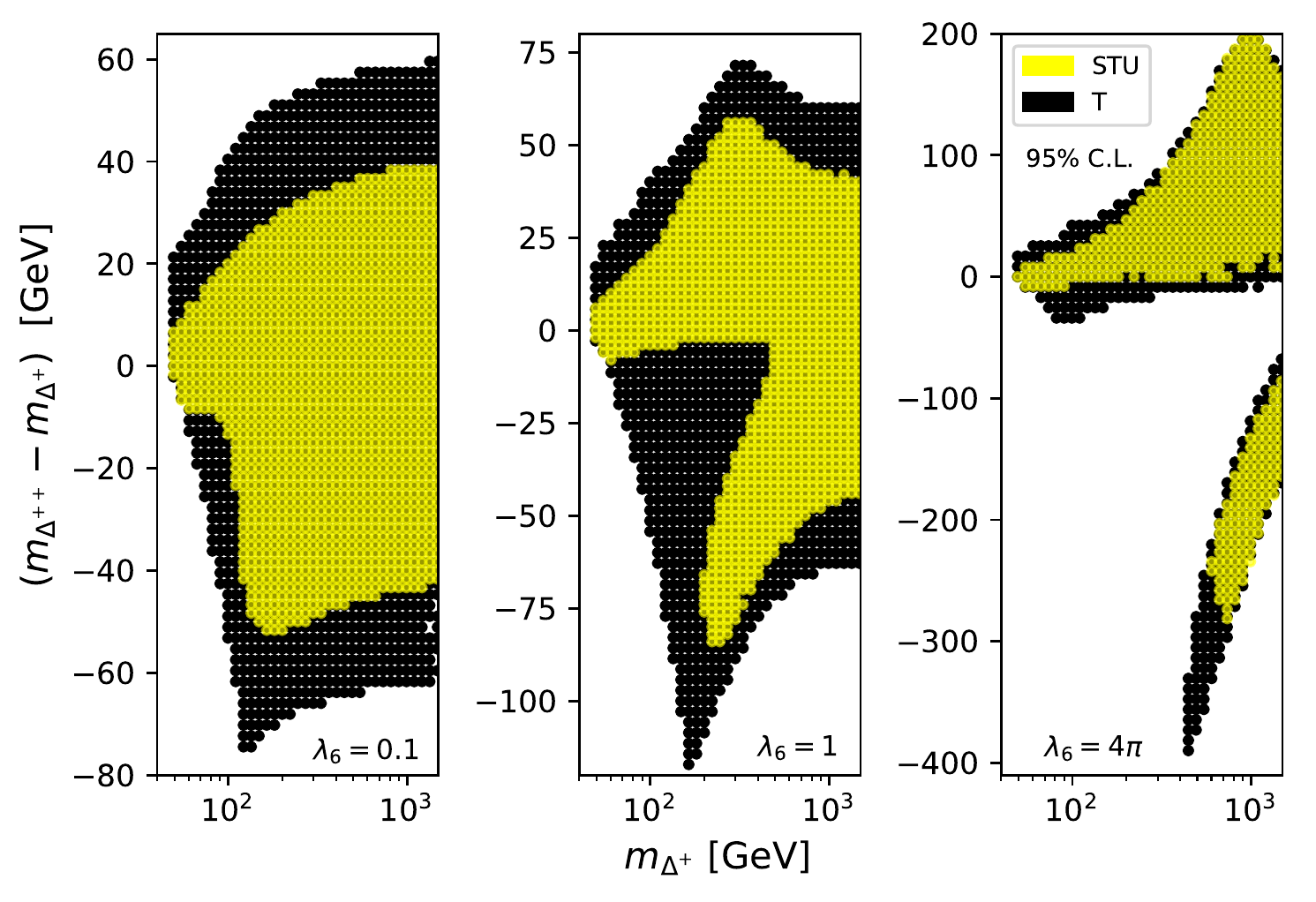}}
\vspace{-2mm}

\caption{\small 
95\% C.L.\ allowed regions in the ($m_{\Delta^{+}}$, $m_{\Delta^{++}} \!-\! m_{\Delta^{+}}$) plane for three different choices of $\lambda_6$ (from left to right, $\lambda_6 = 0.1,\,1,\,4\pi$) in the Inert Triplet model. Yellow regions consider the combination of $S$, $T$ and $U$ electroweak precision constraints, while the allowed regions from considering only the $T$ constraint ($\Delta T \in [-0.21,\,0.27]$) are shown in black.
}
\label{Fig_STU_Triplet}
\end{figure}
The constraints on $S$, $T$ and $U$ restricts the allowed mass splittings among the various BSM states. 
In Figure~\ref{Fig_STU_Triplet} we show allowed model parameter regions under the considered experimental bounds on oblique parameters (see Appendix~\ref{app:A}). Such mass splittings are mainly controlled by the parameters $\tilde{\lambda}_{H\Delta}$ and $\lambda_6$, leading e.g.\ to an allowed splitting $|m_{\Delta^{++}} - m_{\Delta^{+}}| \lesssim 50$ GeV for $\lambda_6 \lesssim 0.1$ (up to $|m_{\Delta^{++}} - m_{\Delta^{+}}| \lesssim 300$ GeV for $\lambda_6 \lesssim 4\pi$) at 95 \% C.L..

\noindent \textsl{\textbf{Dark Matter:}} 
In this scenario DM is a mixture of the scalar $SU(2)_L$ singlet and the ($Y = 1$) triplet, as guaranteed by $\lambda_6 \neq 0$. When this DM state $S_2$ is dominantly singlet-like, the model resembles a Higgs portal singlet DM scenario \cite{Silveira:1985rk,McDonald:1993ex,Burgess:2000yq} (see also~\cite{Feng:2014vea,Athron:2017kgt} and references therein), and the coupling between the $S_2$ state and the SM Higgs is given by
\be
\mathcal{L} \supset - \frac{\lambda_{S_2}}{2}\, S_2^2\, (2\sqrt{2}\, v h + h^2) ,
\ee
with $\lambda_{S_2} = {\rm cos}^2 \alpha \, \lambda_{H\sigma}\,  + {\rm sin}^2 \alpha \, (\lambda_{H\sigma} + \lambda_{H\sigma})/2 - \sqrt{2}\, {\rm sin} \alpha \, {\rm cos} \alpha \, \lambda_6$ being a combination of the various couplings in Eq.~\eqref{eq:LLM}~\cite{Alcaide:2017xoe}. The DM triplet admixture provides additional DM annihilation channels once $m_{S_2} > m_W$. This allows to obtain the correct DM relic density for lower values of $\lambda_{S_2}$ as compared to the singlet Higgs portal model, making it easier to satisfy DM direct detection bounds in the present scenario. We also note that $S_2 - \Delta^{++}$ coannihilations are possible when $\tilde{\lambda}_{H\Delta} > 0$ and $m_{A_0}^2 - m_{S_2}^2 \sim \tilde{\lambda}_{H\Delta} v^2$. It is also worth noting that such a model yields an elegant way to avoid the otherwise extremely stringent direct detection bounds on the DM scattering cross section with nuclei via $Z$-boson exchange which typically rules-out $Y\neq 0$ scalar triplet DM models.\footnote{For an $SU(2)_L$ scalar triplet with $Y = 1$, the mass degeneracy between $\Delta_0$ and $A_0$ states in~\eqref{scalar_triplet} results in a DM scattering cross section with atomic nuclei via $Z$-boson exchange many orders of magnitude above current DM direct detection experimental limits. However, if a mass splitting between $\Delta_0$ and $A_0$ is induced (here due to the singlet-triplet mixing) the $Z$-boson exchange process will become kinematically forbidden and the model could be experimentally viable.}

Besides the tree-level contribution to the DM scattering cross section with nuclei, which is mediated by the Higgs boson and scales like $\lambda_{S_2}^2$, there are important loop-induced DM scattering cross section processes, analogous to those discussed in~\cite{Klasen:2013btp} for the Inert Doublet model. This means that direct detection constraints cannot be completely avoided, even if the freedom in the singlet-triplet mixing\footnote{For the case of singlet-triplet DM models where the scalar triplet instead has $Y = 0$, see, e.g.~\cite{Fischer:2013hwa,Cheung:2013dua}.} $\mathrm{sin}\,\alpha$ allows to obtain the correct DM relic density for $\lambda_{S_2} \to 0$. The loop-induced direct detection signal in this model becomes the same as that of~\cite{Klasen:2013btp}, but now weighted by $\mathrm{sin}^2 \,\alpha$.

Apart from a potential signal in direct DM detection experiments, the model could yield a monochromatic gamma-ray line~\cite{Gustafsson:2007pc,Baumgart:2014saa,Aoki:2015nza,Cirelli:2015bda,Garcia-Cely:2015dda,Garcia-Cely:2016hsk} 
detectable by the FERMI Large Area Telescope or other gamma-ray telescopes. This signal could be particularly strong for a light $\Delta^{++}$ state mediating the DM annihilation into photons. We note that such a signal would be suppressed by $\mathrm{sin}^2\,\alpha$, but at the same time this mixing needs to be sizable in order to avoid current DM direct detection bounds, as explained above. Thus DM direct and indirect detection could be highly complementary to probe this scenario.

\vspace{2mm}
\noindent \textsl{\textbf{LHC Phenomenology:}} 
The collider phenomenology of this model bears some resemblance to the one described above for the {\it Cocktail} model. For $m_{\Delta^{++}} > m_{\Delta^{+}} > m_{A_0}$, the most relevant processes to search for the new scalars at the LHC are those involving the neutral and singly charged states. We note that for sizable values of $\lambda_6$ as needed for neutrino masses, $S_1$ will be significantly heavier than $S_2$. Then, depending on the mass splitting $m_{S_1} - m_{S_2}$ and the singlet-triplet mixing, either the processes $p p \to \Delta^{\pm} S_1$, $p p \to A_0 S_1$, $p p \to \Delta^{\pm} A_0$ or $p p \to \Delta^{\pm} S_2$, $p p \to A_0 S_2$, $p p \to \Delta^{\pm} A_0$ will yield the main avenues for discovery at the LHC. Both lead to multi-lepton signatures similar to those of the Inert Doublet model discussed before.

In contrast, for $m_{A_0} > m_{\Delta^{+}} > m_{\Delta^{++}}$ (e.g. in the coannihilation scenario outlined above) the Drell-Yan processes $p p \to \Delta^{++}\Delta^{--}$ and $p p \to \Delta^{\pm\pm}\Delta^{\mp}$ may yield the dominant probe of this scenario. In all cases the collider bounds are expected similar to those on the Inert Doublet model, only constraining $\mathcal{O}(100)$ GeV masses for the new 
states~\cite{Ilnicka:2015jba,Belyaev:2016lok,Dercks:2018wch,Kalinowski:2019cxe}.

\medskip

To conclude this section, we re-emphasize that already for minimal realizations of {\sl Class~1} neutrino mass scenarios, like the ones discussed above, the phenomenology that emerges is very rich and all such scenarios share many common phenomenological aspects -- ranging from lepton flavour violation to LHC signatures of new charged states. The combination of such various observable aspects serves as a very powerful probe to test these neutrino mass scenarios.

\subsection{Class 2}

The phenomenology of models within this {\sl Class 2} of completions of the $\mathcal{O}_9$ operator in Eq.~\eqref{D9} is in principle more diverse than from those in {\sl Class 1} discussed above. The reason is that, as discussed in section~\ref{sec:EFT}, $\chi$ and $S$ can be both either $SU(2)_L$ singlets or triplets, and either $\chi$ or $S$ can be assigned hypercharge $Y =1$. There are however a few general features worth highlighting:~{\it(i)}   
As opposed to {\sl Class 1} models, there are no lepton flavour violation processes at tree-level in {\sl Class 2} models due to the absence of a $\rho^{++}$ state. The only important lepton flavour violation process is $\mu \to e \gamma$, which occurs at 1-loop.~{\it(ii)} As opposed to {\sl Class 1} models, all the BSM states (including $\chi$ and $S$) are odd under a $\mathbb{Z}_2$ symmetry, which has an important impact on the phenomenology of specific models.~{\it(iii)} Since the models of {\sl Class 2} bear resemblance in particle content (when $S$ carries hypercharge $Y =1$) and neutrino mass matrix structure with the Krauss-Nasri-Trodden radiative neutrino mass model~\cite{Krauss:2002px}, the collider phenomenology and search strategies for the BSM states are also quite similar to those used to probe the Krauss-Nasri-Trodden model and related models~\cite{Ahriche:2013zwa}, in particular regarding multi-leptons~\cite{Guella:2016dwo,Cherigui:2016tbm}.~{\it(iv)} As already mentioned in section~\ref{sec:EFT}, the completion of the operators $\mathcal{O}^\text{2a}_{\scriptscriptstyle{\text{BSM}}}$, $\mathcal{O}^\text{2b}_{\scriptscriptstyle{\text{BSM}}}$ in Eqs.~\eqref{O2a} and~\eqref{O2b} needs in general to occur at tree-level, as otherwise neutrino mass generation happens at 4-loop order, yielding too small neutrino masses to fit neutrino oscillation data.
In the following we discuss several renormalizable models belonging to {\sl Class 2}, corresponding to the states $\chi$ and $S$ being either $SU(2)_L$ singlets or triplets. All these models yield neutrino masses at 3-loops in the presence of DM.

\subsubsection{Model 1: $S$ with hypercharge $Y =1$. $SU(2)_L$ singlets $\chi$ and $S$}
\label{sec:class2_chinesemodel}

This model was introduced in~\cite{Jin:2015cla} (see also~\cite{Geng:2015coa,GR}), considering an extension of the SM by two $SU(2)_L$ singlet fermions $\chi_i \equiv N_{R_i}$ ($i =1,2$) with $Y=0$, and an $SU(2)_L$ singlet, $Y=1$ scalar $S^+$. Here we consider $n > 2$ $SU(2)_L$ singlet fermions (instead of just two) for reasons that we discuss in detail in section~\ref{sec:oscillation_betadecay_class2}. As argued in section~\ref{sec:EFT}, since $S^+$ is an $SU(2)_L$ singlet we need to introduce extra BSM states in order to mediate its interactions with $W$-bosons, and complete the operator $\mathcal{O}^\text{2a}_{\scriptscriptstyle{\text{BSM}}}$ (see Figure~\ref{Fig2} (top-right)). A way to do it at tree-level is to also add to the SM an $SU(2)_L$ scalar triplet with $Y=0$~\cite{Jin:2015cla}
\begin{equation}
\label{Delta_Class2}
\Delta = \left(  \begin{array}{cc}
                  \frac{1}{\sqrt{2}}\Delta_0 & \Delta^+ \\
                  \Delta^- & -\frac{1}{\sqrt{2}}\Delta_0
                 \end{array} \right)\, .
\end{equation}
All the BSM states $\Delta$, $N_{R_i}$ and $S^{+}$ are set to be odd under a $\mathbb{Z}_2$-symmetry. The Lagrangian is then given by 
\begin{eqnarray}
\label{Lag_Singlet}
 \mathcal{L} &=& \frac{1}{2}\, \mathrm{Tr} \left[(D_{\mu}\Delta)^{\dagger}(D^{\mu}\Delta) \right] + (D_{\mu}S)^{*}(D^{\mu}S) + i  
 \overline{N_{R_i}} \slash \hspace{-2mm}\partial N_{R_i}   \nonumber \\
   &-& \frac{1}{2}m_{N_i}\overline{N_{R_i}} N^{c}_{R_i} - g_{ia} \overline{N_{R_i}} \ell^{c}_{R_{a}} S^{+} + \mathrm{h.c.} - V(H,S,\Delta) 
\end{eqnarray}
with $V(H,S,\Delta)$ given by
\begin{eqnarray}
\label{V_Singlet}
\hspace{-0.6cm} V(H,S,\Delta) &=& - \mu_H^2 \left|H\right|^2 + \mu_S^2 \left|S\right|^2 + \mu_{\Delta}^2 \mathrm{Tr} \left[\Delta^2\right]  
 + \lambda_H \left|H\right|^4 + \lambda_S \left|S\right|^4 \nonumber \\
 &+& \lambda_{\Delta} \left(\mathrm{Tr} \left[\Delta^2\right]\right)^2  
  + \lambda_1 \,\left|H\right|^2 \left|S\right|^2 +  \lambda_2 \,\mathrm{Tr}\left[\Delta^2\right] \left|S\right|^2  
 + \lambda_3 \,\mathrm{Tr} \left[\Delta^2\right] \left|H\right|^2 
 \nonumber \\
 &+& \lambda_4 \,H^{\dagger} \Delta \tilde{H} S^{+}  + \mathrm{h.c.}
\end{eqnarray}
The combination of $\lambda_4$ in (\ref{V_Singlet}), $g_{i a}$ and $m_{N_i}$ in (\ref{Lag_Singlet}) breaks lepton number by two units. 
After electroweak symmetry breaking, $\lambda_4$ induces a mixing between $S^+$ and the charged component of the scalar triplet $\Delta^+$. The singlet-triplet mass matrix reads
\begin{equation}
\label{MassMixing_DeltaS}
m^2(\Delta,S) = \left(  \begin{array}{cc}
                  2 \mu_{\Delta}^2 + \lambda_3 v^2 \equiv m^2_{\Delta_0} & \frac{\lambda_4}{2} v^2 \\
                  \frac{\lambda_4}{2} v^2 & \mu_{S}^2 + \frac{\lambda_1}{2} v^2
                 \end{array} \right)
\end{equation}
The singlet-triplet mixing sin$(\beta)$ gives rise to two charged mass eigenstates $H_{1,2}^+$, with
\begin{eqnarray}
\label{MassMixing_DeltaS2}
m^2_{\Delta_0} &=& \mathrm{cos}^2\beta \,m^2_{H_1^+} + \mathrm{sin}^2\beta\, m^2_{H_2^+} \nonumber \\
\lambda_4 v^2 &=& (m^2_{H_2^+} - m^2_{H_1^+})\, \mathrm{sin}\, (2\beta) 
\end{eqnarray}
with $\beta \in [0,\,\pi/2]$ and we restrict ourselves to $\lambda_4 > 0$ so that $m_{H_2^+} > m_{\Delta_0} > m_{H_1^+}$. 

\begin{figure}[t]
\begin{centering}
\includegraphics[width=0.45 \columnwidth]{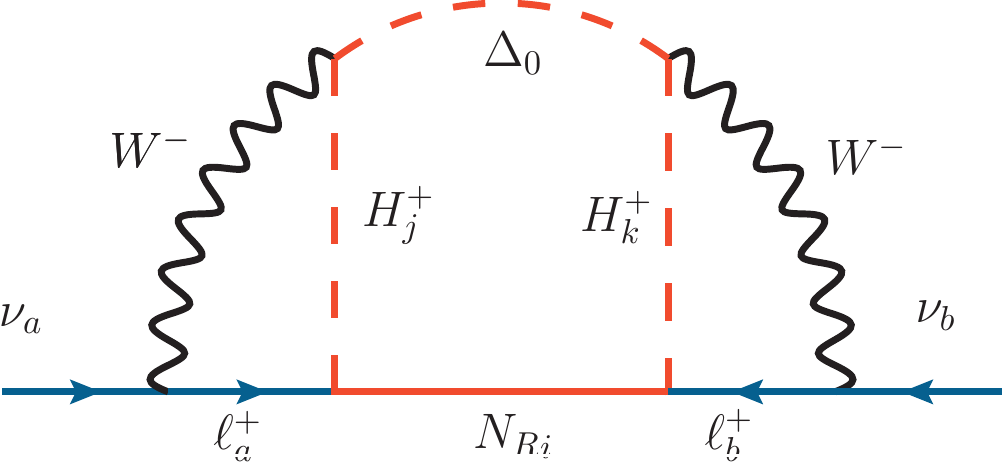}
\hspace{3mm}
\includegraphics[width=0.45 \columnwidth]{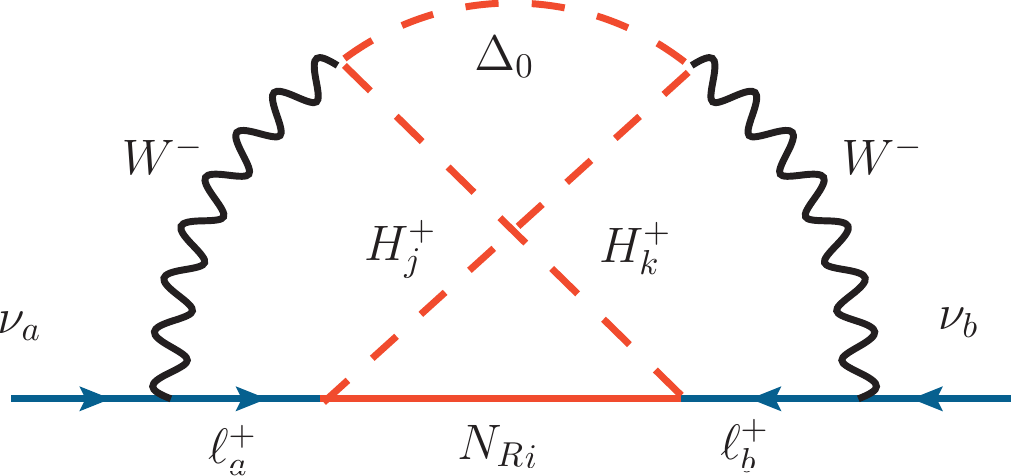}

\vspace{2mm}

\includegraphics[width=0.45 \columnwidth]{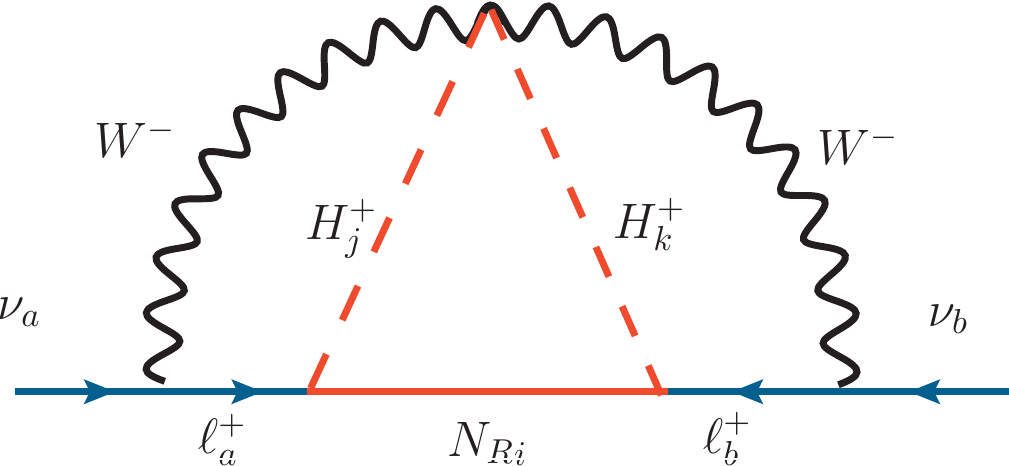}
\par\end{centering}

\caption{\small Three-Loop neutrino mass diagrams for Model 1 ($S$ and $\chi$ being $SU(2)_L$ singlets) within {\sl Class 2}. BSM propagators are depicted in red.}
\label{Fig_Class2_nu}
\end{figure}

\vspace{2mm}

The leading contributions to neutrino masses (shown in Figure~\ref{Fig_Class2_nu}) appear at 3-loops. The neutrino mass matrix is given by (see also~\cite{Jin:2015cla,Geng:2015coa}) 
\begin{eqnarray}
\label{numass_Class2}
m^{\nu}_{ab} =
\frac{m^{\ell}_a \,m^{\ell}_b\,\mathrm{sin}^2 (2\beta)\,(m_{H_1^+}^2-m_{H_2^+}^2)^2}{(16 \pi^2)^3\,\,v^4}
\times \sum_{i=1}^{n}\,\,\sum_{j=1}^{3}\, m_{N_i}\, g_{ia}g_{ib} 
\,I_j(m_{N_i}),
\end{eqnarray}
where $I_j(m_{N_i})$ are 3-loop integrals for the three topologies ($j=1,2,3$) shown in Figure~\ref{Fig_Class2_nu}.

\begin{figure}
\center{\includegraphics[width=0.77 \columnwidth]{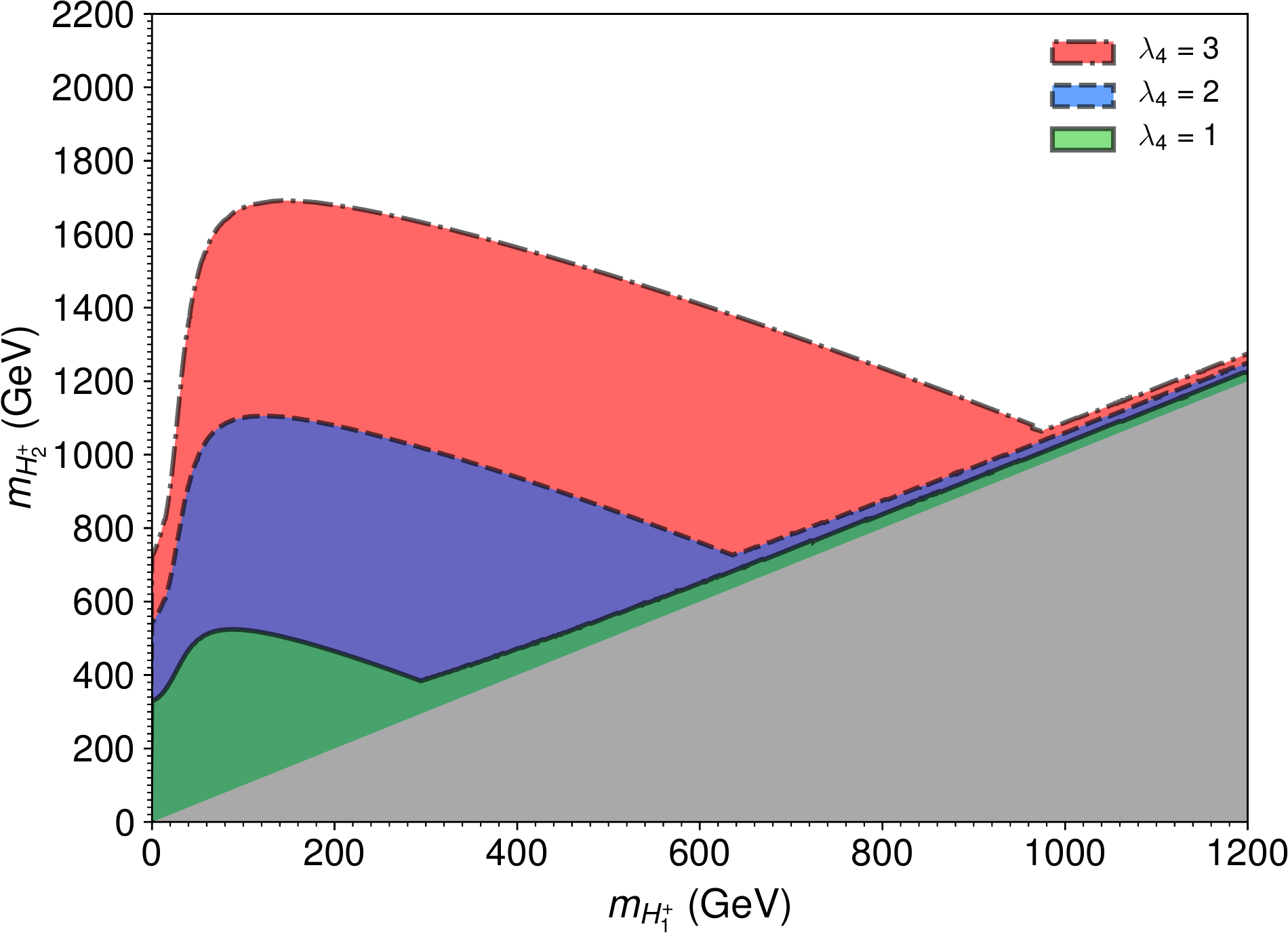}}
\vspace{-2mm}
\caption{\small Excluded region (in colour) in the ($m_{H_1^+}$, $m_{H_2^+}$) plane, for various fixed values of $\lambda_4$, from the combination of bounds on electroweak precision observables ($\Delta T \in [-0.07,\,0.17]$) and Eq.~\eqref{MassMixing_DeltaS2}.}
\label{Fig_Class2_EWPO}
\end{figure}

From Eq.~\eqref{MassMixing_DeltaS2} we observe that there exists an interplay between the generation of sizable neutrino masses, which are proportional to $(\lambda_4 v^2)^2$, and the requirement of satisfying the bounds from electroweak precision observables, in particular from the measurement of the oblique parameter $T$. Its BSM contribution $\Delta T$ is experimentally constrained at 95\% C.L. to the interval (setting in this case $U = 0$) $\Delta T \in [-0.07,\,0.17]$~\cite{PZyla:2020}, and the contribution of the new scalars is given by
\begin{equation}
\Delta T = \frac{1}{4\pi \,\mathrm{sin}^2 \theta_W\, m_W^2} [\mathrm{cos}^2 \beta\, F_{\Delta^0, H_1^+} + \mathrm{sin}^2 \beta \,F_{\Delta^0, H_2^+} - 2\, \mathrm{sin}^2 \beta\, \mathrm{cos}^2 \beta\, F_{H_1^+, H_2^+}]\, ,
\end{equation}
with $\theta_W$ the weak mixing angle and 
\begin{equation}
F_{i,j} = \frac{m_i^2 + m_j^2}{2} - \frac{m_i^2 m_j^2}{m_i^2 - m_j^2} \ln \frac{m_i^2}{m_j^2}\, .
\end{equation}
For a given value of $\lambda_4$, which the neutrino masses depend on directly, there is a minimum value of $m_{H_2^+}^2-m_{H_1^+}^2$ consistent with Eq.~\eqref{MassMixing_DeltaS2}, corresponding to a fixed  $\lambda_4 v^2$ value. We also find that for fixed $\lambda_4$, satisfying the $T$ parameter constraint $\Delta T > -0.07$ imposes a lower bound\footnote{While this may sound counter-intuitive, bear in mind that for fixed $\lambda_4$, a larger mass splitting $m_{H_2^+}-m_{H_1^+}$ results in a smaller singlet-triplet mixing, which ends up balancing the increase in $m_{H_2^+}-m_{H_1^+}$ and overall decreasing the value of $|\Delta T|$.} on $m_{H_2^+}^2-m_{H_1^+}^2$.
This is shown explicitly in Figure~\ref{Fig_Class2_EWPO}.

\medskip

A detailed investigation of the phenomenology of this model and its implications for DM, as well as a comprehensive study of the parameter space for neutrino masses, is left for a forthcoming publication~\cite{GR}. Regarding the phenomenology of the model, we sketch below the main important aspects to consider:

{\it (i)} The collider phenomenology of the model is somewhat similar to that of the Krauss-Nasri-Trodden scenario (as already mentioned above), bearing at the same time much resemblance to {\sl leptophilic} DM scenarios, with the DM candidate (the lightest of the $N_{R_i}$, see below) interacting only with the SM through the SM charged leptons (see e.g.~\cite{Fox:2011fx,Freitas:2014jla} for collider analyses of such scenarios). There is however one important difference in the fact that the present model could give rise to vector boson fusion signatures at colliders, since the kinetic term for the scalar triplet field includes the interaction $g^2 \,W_{\mu}^+ W^{\mu+} \Delta^- \Delta^- + \mathrm{h.c.} \subset \mathrm{Tr} \left[(D_{\mu}\Delta)^{\dagger}(D^{\mu}\Delta) \right]$.

{\it (ii)}  Concerning lepton flavour violation, the present scenario also resembles the Krauss-Nasri-Trodden model (see e.g.~\cite{Cheung:2004xm}), retaining in our case only the new physics contributions proportional to the right-handed charged lepton couplings $g_{ia}$. The $\mu \to e \gamma$ process in particular yields stringent constraints on the allowed parameter space of the model~\cite{GR}.

{\it (iii)}  A viable DM candidate in this model requires the lightest $\mathbb{Z}_2$-odd state to be one of the singlet fermions $N_{R_i}$ (i.e.~the lightest of $N_{R_i}$ must be lighter than the charged state $H_{1}^{\pm}$). We note that the neutral state $\Delta_0$ is always heavier than $H_1^{\pm}$ and thus cannot be the DM. The DM annihilation in the early Universe is mediated by the Yukawa couplings $g_{ia}$ in Eq.~\eqref{Lag_Singlet}, which generically have to be rather large in order to yield neutrino masses compatible with neutrino oscillation data. This leads to a suppressed DM relic density unless $m_{N_{R_i}}$ is correspondingly large, and it is not completely clear if such a setup is phenomenologically viable (e.g.~allowing for neutrino masses of the correct size).

\subsubsection{Model 2: $S$ with hypercharge $Y =1$. $SU(2)_L$ triplets $\chi$ and $S$}

We now introduce a novel UV completion to the operator $\mathcal{O}^\text{2a}_{\scriptscriptstyle{\text{BSM}}}$ for which 
the fermions $\chi_i$ and scalar $S$ in Eq.~\eqref{D9UV_2} are $SU(2)_L$ triplets. We extend the SM by $n > 2$ $SU(2)_L$ triplet fermions $\chi_i \equiv \Sigma_i$ with $Y = 0$ and an $SU(2)_L$ triplet scalar $S \equiv \Delta$ with hypercharge $Y = 1$, and will now use the labels $\Sigma$ and $\Delta$ for our fermion and scalar triplets. Since the state $\Delta$ couples to the SM gauge bosons, we could in principle hope to generate neutrino masses only through the presence of these two BSM fields.
Imposing that both fields are odd under a $\mathbb{Z}_2$-symmetry, the Lagrangian is
\begin{eqnarray}
\label{Lag_Triplet}
 \mathcal{L} &=& \frac{1}{2}\, \mathrm{Tr} \left[(D_{\mu}\Delta)^{\dagger}(D^{\mu}\Delta) 
 \right] + i \overline{\Sigma_{R_i}} \slash \hspace{-2mm}D \Sigma_{R_i} \nonumber \\
   &-& \frac{1}{2}m_{\Sigma_i}\overline{\Sigma_{R_i}} \Sigma^{c}_{R_i} - g_{ia} 
   \overline{\Sigma_{R_i}} \ell^{c}_{R_{a}} \Delta + \mathrm{h.c.} - V(H,\Delta) 
\end{eqnarray}
with $V(H,\Delta)$ depending on even powers of $H$ and $\Delta$.
Since $\Delta$ has hypercharge $Y=1$, the Lagrangian (\ref{Lag_Triplet}) turns out not to violate lepton number, thus failing to generate neutrino masses. In order to violate lepton number, we further introduce a real singlet scalar field $\sigma$. This introduces (among others) the following additional terms to the 
Lagrangian 
\begin{equation}
m^2_{\sigma}\sigma^2 + \lambda_5 \sigma H^{\dagger} \Delta \tilde{H} + \mathrm{h.c.},
\end{equation}
which together with the terms in \eqref{Lag_Triplet} break lepton number by two units. 

Upon electroweak symmetry breaking, the CP-even neutral component of $\Delta$, which we label $\Delta_0$, mixes with the singlet scalar $\sigma$ due to the $\lambda_5$ term. When the lightest $\mathbb{Z}_2$-odd neutral scalar is lighter than the fermion triplet $\Sigma$, the $\Delta_0-\sigma$ mixing makes this scenario viable from the point of view of DM, since DM from a pure inert triplet with $Y=1$ is extremely constrained by DM direct detection experiments, whereas the mixing provides a mass splitting between the states $\Delta_0$ and $A_0$ (the CP-odd neutral component of $\Delta$) within the scalar triplet, thus avoiding the direct detection bounds (recall the discussion in section~\ref{sec:lollipop}). Alternatively, when the fermion triplet is lighter than all the $\mathbb{Z}_2$-odd scalars, the neutral component of the fermion triplet $\Sigma_0$ yields a viable DM candidate, being in fact a particular realization of the Minimal DM scenario~\cite{Cirelli:2005uq}. The DM mass $m_{\Sigma_0}$ yielding the correct DM relic density is in this case $m_{\Sigma_0} \sim 2.4$ TeV~\cite{Cirelli:2005uq}.

\subsubsection{Models 3: $\chi$ with hypercharge $Y =1$} 

As discussed in section~\ref{sec:EFT}, it is also possible to generate neutrino masses via the operator $\mathcal{O}^\text{2b}_{\scriptscriptstyle{\text{BSM}}}$ in~\eqref{O2b}, which together with the interaction~\eqref{D9UV_2} and the mass term $m_s^2 S^2$ break lepton number by two units. In this case, the state $\chi$ has hypercharge $Y =1$ and $S$ has hypercharge $Y =0$, and they can be either $SU(2)_L$ singlets or triplets. 
In both cases, new physics is required to allow the state $\chi$ to couple to the SM $SU(2)_L$ gauge bosons in an appropriate way, in order to yield a renormalizable completion of $\mathcal{O}^\text{2b}_{\scriptscriptstyle{\text{BSM}}}$.

As we are not aware of any specific neutrino mass model of this type in the literature, we outline in the following two possible setups yielding a renormalizable completion of the $\mathcal{O}^\text{2b}_{\scriptscriptstyle{\text{BSM}}}$ operator: one with $\chi$ and $S$ being $SU(2)_L$ singlets, and then one with $\chi$ and $S$ being $SU(2)_L$ triplets. In both setups we consider $\chi$ and $S$ to be odd under a $\mathbb{Z}_2$-symmetry.

\vspace{2mm}

When $\chi$ and $S$ are $SU(2)_L$ singlets, the coupling of $\chi$ to the SM gauge bosons may be obtained by introducing an $SU(2)_L$ doublet vector-like\footnote{In order to guarantee that the model remains gauge anomaly-free.} lepton $\Psi$ with hypercharge $Y = 1/2$ and a Yukawa interaction with $\chi$ 
\be
\label{O8_b_model_extraterm}
Y_\chi\,\overline{\Psi} \tilde{H} \chi + \mathrm{h.c.}\,,
\ee
with $H$ the SM Higgs doublet (and $\tilde{H} = i \sigma_2 H^*$).
Upon electroweak symmetry breaking, the interaction \eqref{O8_b_model_extraterm} leads to a mixing between $\chi$ and the charged component of $\Psi$, which then induces the desired coupling between $\chi$ and the SM gauge bosons. 
The term \eqref{O8_b_model_extraterm} is however not enough to yield lepton number violation in combination with the interaction \eqref{D9UV_2} and the mass term $m_s^2 S^2$, so some additional ingredient would be required to generate a neutrino mass. We can further add a neutral singlet Majorana fermion $\psi$ with the Lagrangian terms
\be
\label{O8_b_model_extraterm2}
Y_\psi\,\overline{\Psi} H \psi + m_{\psi}\,\bar{\psi}^c\psi + \mathrm{h.c.}  \,.
\ee
The first term yields a mixing between the neutral component of $\Psi$ and the singlet state $\psi$. The combination of~\eqref{O8_b_model_extraterm} and~\eqref{O8_b_model_extraterm2} allows to UV complete the operator $\mathcal{O}^\text{2b}_{\scriptscriptstyle{\text{BSM}}}$ and give rise to lepton number violation together with \eqref{D9UV_2} and the mass term $m_s^2 S^2$. Yet, this is achieved at the expense of introducing quite a number of BSM fields.

\vspace{2mm}

When $\chi$ and $S$ are instead $SU(2)_L$ triplets, a coupling of $\chi$ to the SM gauge bosons is granted, but it does not directly allow to complete the $\mathcal{O}^\text{2b}_{\scriptscriptstyle{\text{BSM}}}$ in a renormalizable manner, as the $\chi$ gauge interactions do not lead to lepton number violation.
In order to achieve this via  
direct mixing\footnote{The absence of such a mixing (e.g.\ lepton number violation via interactions between $\chi$ and other BSM fields, which do not result in a mixing with them) would lead to neutrino masses generated at $4$-loop order or beyond.} with other BSM states, a possibility is again to introduce a vector-like fermion doublet\footnote{See~\cite{Dedes:2014hga} for a related implementation of this mechanism for doublet-triplet fermionic DM with hypercharge $Y = 0$.} $\Psi$ with hypercharge Y = 1/2 and neutral singlet Majorana fermion $\psi$ with the Lagrangian terms
\be
\label{O8_b_model_extraterm3}
Y_\chi\,\overline{\chi} (H \vec{\sigma} \Psi) + Y_\psi \,\overline{\Psi} H \psi  + m_{\psi}\,\bar{\psi}^c\psi + \mathrm{h.c.}\,,
\ee
with $\vec{\sigma}$ the Pauli matrices vector and $H \vec{\sigma} \Psi$ transforming as an $SU(2)_L$ triplet fermion with hypercharge $Y = 1$. After electroweak symmetry breaking the Majorana singlet fermion $\psi$ mixes with the neutral components of $\Psi$ and $\chi$, allowing for lepton number violation and neutrino mass generation at three-loops. Here also the DM candidate may be the neutral component of the $SU(2)_L$ triplet scalar $S$ (with hypercharge $Y=0$), with a phenomenology similar to that of minimal DM~\cite{Cirelli:2005uq}. As is clear from the discussion in this section, these models are rather cumbersome, and we do not explore them further in this work.

\medskip

In summary, we find that {\sl Class 2} neutrino mass scenarios have been barely explored in the literature, and we leave a more detailed study of specific models for future work~\cite{GR}. Similarly to {\sl Class 1}, their phenomenology is expected to be very rich but still somewhat different, in particular regarding lepton flavour violation due to the absence of the BSM state $\rho^{++}$ in {\sl Class 2} scenarios. In addition, we show in the next section that there are also important differences between {\sl Class 1} and {\sl Class 2} regarding neutrinoless double $\beta$-decay signatures and their interplay with the pattern of neutrino masses and mixings.

\section{Neutrino mixing and neutrinoless double $\beta$-decay}
\label{sec:Constraints}

We now discuss the interplay between the requirements of fitting the oscillation data for neutrino masses and mixings, and satisfying the bounds from neutrinoless double $\beta$-decay experiments. As we will see below, the combination of these aspects has important consequences for renormalizable completions of the effective operator~$\mathcal{O}_9$ (this has been studied in some depth for {\sl Class 1} models in~\cite{Gustafsson:2014vpa}).  

\vspace{2mm}

For the case of Majorana neutrinos, a parametrization of their mass matrix, in the basis where charged current interactions are flavour-diagonal and the charged leptons $e, \mu, \tau$ are simultaneously mass eigenstates, reads
\be
\label{UPMNS}
m^{\nu} = U^{T} \,m^{\nu}_{D} \,U \quad \mathrm{with} \quad m^{\nu}_{D} = 
\mathrm{Diag}\left(m_1,m_2,m_3\right).
\ee
Here $m_{1,2,3}$ are the masses of the three light neutrinos and $U^{T}$ is the PMNS matrix \cite{Pontecorvo:1957qd,Maki:1962mu}, given in terms of three mixing angles $\theta_{12}$, $\theta_{23}$, $\theta_{13}$ and three phases, a CP phase $\delta$ and two Majorana phases $\alpha_1$ and $\alpha_2$:
\bea
\label{UPMNS2}
U = \mathrm{Diag}\left(1,e^{i\alpha_{1}},e^{i(\alpha_{2}+\delta)}\right) \times \quad \quad \quad 
\quad \quad \quad 
\\
\hspace{-0.3cm}
\left(
\begin{array}{lcc}
c_{13}c_{12} & -c_{23}s_{12}- s_{23}c_{12}s_{13}e^{i\delta} & \;\;s_{23}s_{12}- c_{23}c_{12}s_{13}e^{i\delta} 
\\    
c_{13}s_{12}  & \;\;c_{23}c_{12}- s_{23}s_{12}s_{13}e^{i\delta} &  -s_{23}c_{12}- c_{23}s_{12}s_{13}e^{i\delta} 
\\
s_{13}e^{-i\delta} & s_{23}c_{13}  & c_{23}c_{13}
\end{array}
\right)\!\!,\nonumber
\eea
where $s_{ij} \equiv \mathrm{sin}(\theta_{ij})$ and $c_{ij} \equiv \mathrm{cos}(\theta_{ij})$. 
A global fit to the data from different neutrino oscillation experiments 
gives~\cite{Esteban:2018azc} $\Delta m^2_{21} \equiv m^2_{2} - m^2_{1} 
= 7.39^{+0.21}_{-0.20}\times 10^{-5} \mathrm{eV}^2$, $\left|\Delta m^2_{31}\right| \equiv 
\left| m^2_{3} - m^2_{1} \right|= 2.525^{+0.033}_{-0.031}\times 10^{-3} \mathrm{eV}^2$  
($2.512^{+0.031}_{-0.034}\times 10^{-3} \mathrm{eV}^2$) for $\Delta m^2_{31} > 0$ ($\Delta m^2_{31}<0$),  $s_{12}^2 = 0.310^{+0.013}_{-0.012}$, $s_{13}^2 = 0.02240^{+0.00065}_{-0.00066}$ 
and $s_{23}^2 = 0.582^{+0.015}_{-0.019}$ (the atmospheric angle solution in the first octant, $s_{23}^2 \simeq 0.46$, is currently disfavoured by a bit less than $2\sigma$). Neutrino oscillation experiments are still not fully sensitive to the sign of $\Delta m^2_{31}$ which results in two possible mass orderings in the neutrino sector. These are known as normal ordering (NO) and inverted ordering (IO) and are characterized by
\be
\begin{array}{l}
\Delta m^2_{31} > 0 \quad \rightarrow \quad  m_1 < m_2 < m_3 \quad \mathrm{(NO)}\\
\Delta m^2_{31} < 0 \quad \rightarrow \quad   m_3 < m_1 < m_2 \quad \mathrm{(IO)\,,}
\end{array}
\ee
with current oscillation data favouring NO over IO at the $2\sigma-3\sigma$ level.

\vspace{2mm}

\begin{figure}[t!]
\center{
\includegraphics[width=0.25 \columnwidth]{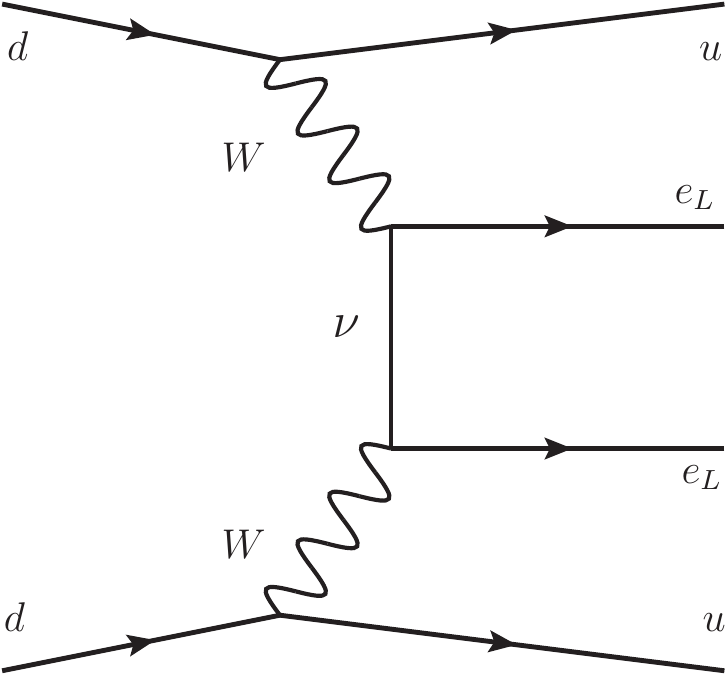}\hspace{2cm} 
\includegraphics[width=0.25\columnwidth]{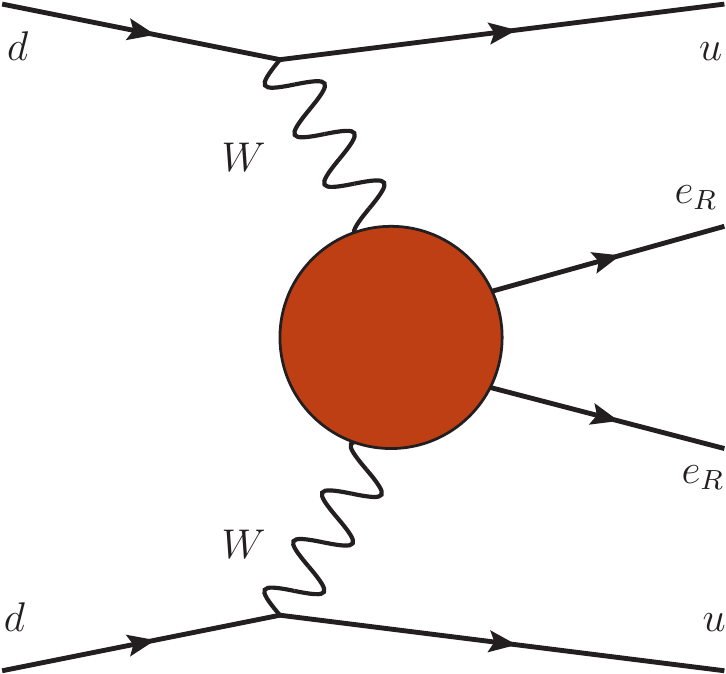}}
\caption{Contributions to the neutrinoless double $\beta$-decay process. Left: light neutrino exchange (long-distance contribution), proportional to $m^{\nu}_{ee}$. Right:~short-distance contribution proportional to $\epsilon_3$. The solid red blob corresponds to the $\mathcal{O}_9$ effective operator.}
\label{fig:tree_level_short_D_Nu0bb}
\end{figure}

Now we turn to look at the neutrinoless double $\beta$-decay probe of Majorana neutrinos. Here we distinguish between two contributions to the neutrinoless double $\beta$-decay process (as shown i Figure~\ref{fig:tree_level_short_D_Nu0bb}): a long-distance contribution involving light Majorana neutrinos (left panel) and a short-distance contribution that directly involves the lepton number violating effective operator that is responsible for generating neutrino masses, in our case the $\mathcal{O}_9$ operator (right panel).

\vspace{2mm}

Both the long- and short-distance contributions to neutrinoless double $\beta$-decay are present in general, and which one is dominant depends on the specific type of lepton number violating new physics considered. 
For the scenarios studied in this work, the operator $\mathcal{O}_9$ leads to a neutrino mass matrix at two loops or higher (recall Figure~\ref{Fig1}), with an unavoidably small $m^\nu_{ee}$ element, $m^\nu_{ee} \ll 10^{-4}$ eV. Thus, the contribution to the neutrinoless double $\beta$-decay amplitude due to a light Majorana neutrino propagator, being proportional to $m^\nu_{ee}/ p^2$ (with $p^2 \sim (100\;\text{MeV})^2$), is extremely small. However, the $\mathcal{O}_9$ operator also gives rise to an effective tree-level short-distance contribution from the diagram shown in the right panel of Figure~\ref{fig:tree_level_short_D_Nu0bb}, which does not suffer from the extra loop and $(m^{\ell}_e/v)^2$ suppression that affects the $m^{\nu}_{ee}$ neutrino mass entry, will largely dominate the neutrinoless double $\beta$-decay amplitude~\cite{delAguila:2011gr,Gustafsson:2014vpa}.

The emerging effective six-fermion contact interaction from the short-distance contribution corresponding to the right diagram in Figure~\ref{fig:tree_level_short_D_Nu0bb} can be written as (see Refs.~\cite{Pas:2000vn,Bergstrom:2011dt})
\be
\label{Lbetabeta}
\mathcal{L}_{0\nu\beta\beta} =  \frac{G_F^2}{2\, m_p}\, \epsilon_3 \, J^{\mu}\,J_{\mu}\, \bar{e}(1-\gamma_5)e^c.
\ee 
where $J^{\mu} = \bar{u} \gamma^{\mu} (1-\gamma_5) d$ is the vector-axial hadronic current and
\be
\label{E3_1}
\epsilon_3 = - 2\,m_p\, \mathcal{A}^{\mathrm{SD}}_{0\nu\beta\beta}\, ,
\ee 
with $m_p$ the proton mass and $\mathcal{A}^{\mathrm{SD}}_{0\nu\beta\beta}$ the total Feynman amplitude for the short-distance contribution to neutrinoless double $\beta$-decay. 
The nuclear isotope half-life time $T^{0\nu\beta\beta}_{1/2}$ from neutrinoless double $\beta$-decays due to light neutrino exchange and 1-loop short-distance (SD) interactions are, respectively (see e.g.~\cite{Horoi:2017gmj}),
\be
\label{HalfLife}
\left[ T^{0\nu\beta\beta}_{1/2} \right]^{-1}  =  g_A^4 \,G_{01}\, \frac{|m^{\nu}_{ee}|^2}{(m^{\ell}_e)^2}\, |\mathcal{M}^{\mathrm{\nu}}|^2
\quad \quad \mathrm{and} \quad \quad 
\left[ T^{0\nu\beta\beta}_{1/2} \right]^{-1} =  g_A^4\, G_{01}\, |\epsilon_3|^2\, |\mathcal{M}^{\mathrm{SD}}|^2.
\ee 
Here $g_A = 1.27$ is the axial vector coupling constant and $G_{01}$ is a nuclear isotope specific phase-space factor while $\mathcal{M}^{\nu}$ (light neutrino exchange) and $\mathcal{M}^{\mathrm{SD}}$ (short-distance) are nuclear matrix elements (NMEs) for the specific isotope. Values on these parameters, together with current experimental limits and future projections for $T^{0\nu\beta\beta}_{1/2}$ for different isotopes can be found in Table~\ref{tab:1} (for a comprehensive review, see~\cite{Dolinski:2019nrj}).

\vspace{1mm}

The need to fit the neutrino mass and mixing data from oscillation experiments combined with the bounds from neutrinoless double $\beta$-decay affect models within {\sl Class 1} and {\sl Class 2} in different ways, as we discuss in the following.

\begin{table}[t]
\begin{tabular}{rcc c c c}
\hline\hline \\[-4mm]
& $G_{01}\,\, (10^{-14}\, \mathrm{yr}^{-1})$ & $\left|\mathcal{M}^{\nu} \right|$ &  $\left|\mathcal{M}^{\mathrm{SD}} \right|$  & $T^{0\nu\beta\beta}_{1/2}$ limit ($\times 10^{26}$~yr)  & future ($\times 10^{26}$~yr)\\[0.5mm]
\hline
$\,^{76}\mathrm{Ge}$ & 0.22  & 2.52 & 208.4 & 0.8~\cite{Agostini:2018tnm} & 
10-100~\cite{Abgrall:2017syy}\cr
$\,^{136}\mathrm{Xe}$ & 1.45 & 1.74 & 106.4 & 1.07~\cite{KamLAND-Zen:2016pfg} & 20-100~\cite{Dekens:2020ttz,Albert:2017hjq}\cr
$\,^{130}\mathrm{Te}$  & 1.41  & 2.25  & 192.8 & 0.15~\cite{Alduino:2017ehq} & 
$>$ 10~\cite{Artusa:2016mat,CUPIDInterestGroup:2019inu}\cr
\hline\hline
\end{tabular}
\caption{\small Values of the phase-space factors $G_{01}$ and nuclear matrix elements $\left|\mathcal{M}^{\nu} \right|$ and $\left|\mathcal{M}^{\mathrm{SD}} \right|$ for different nuclear isotopes. The values of $G_{01}$ are obtained from~\cite{Horoi:2017gmj,Neacsu:2015uja} (see also~\cite{Suhonen:1998ck,Stefanik:2015twa}). 
The nuclear matrix elements (NMEs) $\left|\mathcal{M}^{\nu} \right|$ are the averaged best estimates from Ref.~\cite{GomezCadenas:2010gs}. The NMEs $\left|\mathcal{M}^{\mathrm{SD}} \right|$ are the values given in Ref.~\cite{Deppisch:2012nb} from their computation using the proton-neutron quasiparticle random phase approximation (pn-QRPA) approach. For $\,^{76}\mathrm{Ge}$ and $\,^{136}\mathrm{Xe}$ they are found to agree to better than 30\% with the NME computation from~\cite{Horoi:2017gmj} using the interacting shell model (ISM), while for $\,^{130}\mathrm{Te}$ there is roughly a factor two mismatch with ISM. The last two columns show the current best limit on $T^{0\nu\beta\beta}_{1/2}$ for each isotope, as well as the approximate reach of planned experiments for this decade.}
\label{tab:1}
\end{table}

\subsection{Class 1}

As presented in section~\ref{sec:EFT}, in {\sl Class~1} scenarios the electron-electron entry of the Majorana neutrino mass matrix, is given by (recall Eq.~\eqref{NuMassClass1})
\be
\label{NuMassClass1_ee}
m^{\nu}_{ee} = \frac{C_{ee}}{(16\pi^2)^{L+2}}\, 
\left(\frac{m^{\ell}_e}{v}\right)^2 \, \frac{v^2}{\Lambda} \, , 
\ee
is very suppressed, $m^\nu_{ee} \ll 10^{-4}$ eV, irrespectively of the value of the Yukawa coupling $C_{ee}$ (as long as it remains perturbative). This approximate neutrino mass texture leads to correlations among neutrino mixing parameters (see~\cite{Gustafsson:2014vpa} for a detailed discussion), via  
\be
\label{mee}
m^{\nu}_{ee} \equiv c_{13}^2 \left(m_1 c_{12}^2 + e^{2i \alpha_1}  m_2  s_{12}^2 \right) + 
 e^{2i \alpha_2}  m_3  s_{13}^2 \sim 0
\ee
The same suppression partially affects also the $m^{\nu}_{e\mu}$ entry, proportional to $C_{e\mu}$ and yet to a much lesser extent the $m^{\nu}_{e\tau}$ entry proportional to $C_{e\tau}$. Altogether this leads
to several predictions for the ranges of neutrino oscillation parameters, including a NO for neutrino masses, the lightest neutrino mass in the $\sim$~meV range and a specific correlation between the values of $\theta_{13}$ and 
$\theta_{23}$ (see~\cite{Gustafsson:2014vpa} for details). 

\vspace{1mm}

In contrast, for neutrinoless double $\beta$-decay the tree-level short-distance contribution is induced directly by the operator $\mathcal{O}^{1}_{\rm BSM}$ (which is the core of $\mathcal{O}_9$ for {\sl Class 1} models), and does not carry the extra two loops and $(m^{\ell}_e/v)^2$ suppression that is needed to generate $m^{\nu}_{ee}$. Instead the neutrinoless double $\beta$-decay rates are directly proportional to $C_{ee}$. From LFV searches, $C_{ee}$ is  constrained to not be too large due to the processes $\rho^{++}$ otherwise induce (discussed in section~\ref{sec:class.1}).
Yet, we stress that neutrino oscillation data do not impose restriction on any (perturbative) values of $C_{ee}$. 

As a consequence, there is 
no correlation between the constraints imposed on the parameters of neutrino mass models of {\sl Class 1} by neutrino oscillation data, and the predictions for neutrinoless double $\beta$-decay in this models. Experimental bounds on $\epsilon_3$ can thus be trivially satisfied since $C_{ee}$ is in essence a free parameter, whose value does not affect neutrino masses and mixings in these models.

\subsection{Class 2}
\label{sec:oscillation_betadecay_class2}

For {\sl Class 2} models, the neutrino mass matrix can be generically written as (recall Eq.~\eqref{NuMassClass2}) 
\be
\label{NuMassClass2_ab}
m^{\nu}_{ab} = \sum_{i=1}^n\frac{g_{ia} g_{ib}}{(16\pi^2)^{L+3}}\, \frac{m^{\ell}_a\, m^{\ell}_b}{v^2} \, \frac{v^2}{\Lambda'}, 
\ee
with $i = 1,...,n$ the number of BSM fermions $\chi_i$ in the model. We show in the following that these models face some difficulty in fitting the neutrino oscillation data. 
For radiative neutrino mass models with BSM fermions $\chi_i$ where the flavour structure of the neutrino mass matrix depends on a product of Yukawa couplings, $m^{\nu}_{ab} \propto g_{ia} g_{ib}$, at least $n = 2$ is required to fit neutrino oscillation data (as shown in~\cite{Cheung:2004xm} for the particular case of the Krauss-Nasri-Trodden model). Yet, we demonstrate below that $n = 2$ is not sufficient for {\sl Class 2} models to fit oscillation data.

\vspace{1mm}

As for models of {\sl Class 1}, here the $m^\nu_{ee}$ entry of the neutrino mass matrix has to be (for perturbative couplings $g_{ie}$) very small, $m^\nu_{ee} \ll 10^{-4}$ eV, due to the $(m_e^{\ell}/v)^2$ factor combined with the 3-loop suppression in~\eqref{NuMassClass2_ab}. If $n = 2$, then the neutrino mass matrix has  vanishing determinant, $\mathrm{Det}(m^{\nu}) = 0$, since
\begin{equation}
\label{numass_two}
m^{\nu}_{ab} \propto \left(  \begin{array}{ccc}
                  g_{1e} & g_{2e} & 0 \\
                  g_{1\mu} & g_{2\mu} & 0 \\
                  g_{1\tau} & g_{2\tau} & 0
                 \end{array} \right) 
                 \,\left(  \begin{array}{ccc}
                  g_{1e} & g_{1\mu} & g_{1\tau} \\
                  g_{2e} & g_{2\mu} & g_{2\tau} \\
                  0 & 0 & 0 \\
                 \end{array} \right)
\end{equation}
and as consequence the lightest neutrino is massless. We can then use the dependence of $m_{ee}^{\nu}$ on the neutrino masses and mixings (see Eq.~\eqref{mee}) to obtain a lower bound on $|m^{\nu}_{ee}|$ for NO and IO
\begin{eqnarray}
 |m^{\nu}_{ee}| > \left|\sqrt{\Delta m_{31}^{2}} s^2_{13} - \sqrt{\Delta m_{12}^{2}} s^2_{12} c^2_{13}\right| \quad \quad (\mathrm{NO}) \nonumber\\
 \label{boundmee}\\
 |m^{\nu}_{ee}| > c^2_{13}\,\sqrt{\Delta m_{13}^{2}}\, \left| s^2_{12} - c_{12}^2 \sqrt{1- \frac{\Delta m_{21}^{2}}{\Delta m_{13}^{2}} } \right|  \quad \quad (\mathrm{IO}) \nonumber
\end{eqnarray}
which altogether yield $|m^\nu_{ee}| \gtrsim 0.001$~eV, clearly incompatible with 
the strong $m^\nu_{ee}$ suppression discussed above. We note that for the only specific neutrino mass model of {\sl Class 2} existing in the literature~\cite{Jin:2015cla} (see section~\ref{sec:class2_chinesemodel}), $n = 2$ was considered, leading to inconsistently large couplings (well beyond the $4\pi$ perturbativity limit) for the model, as pointed out in~\cite{Geng:2015coa}.   

The above problem is however solved for $n \geq 3$, since in this case the lightest neutrino need not be massless and the lower bounds~\eqref{boundmee} do not apply. Yet, for $m^{\nu}_{ee} \ll 10^{-4}$ eV, either $m^{\nu}_{e\tau} \sim m^{\nu}_{\mu\mu}\sim m^{\nu}_{\tau\tau} \sim \sqrt{|\Delta m_{31}^{2}|}$ or  $m^{\nu}_{e\mu} \sim m^{\nu}_{\mu\mu}\sim m^{\nu}_{\tau\tau} \sim \sqrt{|\Delta m_{31}^{2}|}$ would be approximately required to fit neutrino oscillation data~\cite{Gustafsson:2014vpa}. We focus in the following discussion on $m^{\nu}_{e\tau} \sim \sqrt{|\Delta m_{31}^{2}|}$, bearing in mind that the discussion of the latter option (with a sizable $m^{\nu}_{e\mu}\sim \sqrt{|\Delta m_{31}^{2}|}$) is analogous in essence but more difficult to be realized phenomenologically. 
Then, since 
\be
m^{\nu}_{ab} \propto \sum_{i=1}^n g_{ia}\, g_{ib}\, \frac{m^{\ell}_a\, m^{\ell}_b}{v} \, ,
\ee
if we assume no cancellation among the $g_{ia}\, g_{ib}$ contributions from different $\chi_i$ states, $m^{\nu}_{e\tau} \sim m^{\nu}_{\tau\tau}$ leads to $g_{ie} \sim g_{i\tau} \times (m^{\ell}_{\tau} / m^{\ell}_{e}) \gg g_{i\tau}$ and then we would also have $ m^{\nu}_{\tau\tau} \sim m^{\nu}_{ee} \ll 10^{-4}$~eV, again revealing an impossibility to fit neutrino oscillation data. Thus, for $n \geq 3$ there have to exist cancellations among the contributions to (some of) the $m^{\nu}_{ab}$ entries from different $\chi_i$ states. This can be illustrated using the model from {\sl Class 2a} introduced in section~\ref{sec:class2_chinesemodel} (a detailed analysis of this model is left for the future~\cite{GR}), which contains a charged $SU(2)_{L}$ singlet scalar $S^{+}$ and a set of neutral singlet fermions $\chi_i \equiv N_{R_i}$ (with $i = 1,..., n$; $n \geq 3$) as BSM states.
The neutrino mass matrix is given in this case by (recall Eq.~\ref{numass_Class2}) 
\begin{eqnarray}
\label{numass_Class2_sec4}
m^{\nu}_{ab} = \lambda_4^2 \times
\frac{m^{\ell}_a \,m^{\ell}_b}{(16 \pi^2)^3}
\, \sum_{i=1}^{n}\, g_{ia}g_{ib}\, m_{N_i}\, \sum_{j=1}^{3}
\,I_j(m_{N_i}) = 
\frac{\lambda_4^2\, m^{\ell}_a \,m^{\ell}_b}{(16 \pi^2)^3}
\, \sum_{i}\, g_{ia}g_{ib} \,m_{N_i} \,I_i
\end{eqnarray}
with $I_j(m_{N_i})$ are 3-loop integrals for the three topologies ($j=1,2,3$) shown in Figure~\ref{Fig_Class2_nu}, and $\lambda_4^2$ a dimensionless parameter of the model (see section~\ref{sec:class2_chinesemodel}). Neutrino masses of the right size require $\mathcal{O}(1)$ Yukawa couplings and no cancellation among different $g_{ie}g_{i\tau}$ terms in the $m^{\nu}_{e\tau}$ entry. Then, fitting the observed neutrino mixing pattern from oscillations demands
\be
m^{\nu}_{e\tau} \sim m^{\nu}_{\tau\tau} \, ,
\ee
which leads to the hierarchy
\be
\sum_{i}\, g_{ie}g_{i\tau} \,m_{N_i} \,I_i \gg \sum_{i}\, g_{i\tau}g_{i\tau} \,m_{N_i} \,I_i
\ee
for $g_{ie} \sim g_{i\tau} \sim 1$. This clearly implies in general a strong cancellation among different $g_{i\tau}g_{i\tau}$ contributions to $m^{\nu}_{\tau\tau}$.

\begin{figure}[t!]
\center{\includegraphics[width=0.28 \columnwidth]{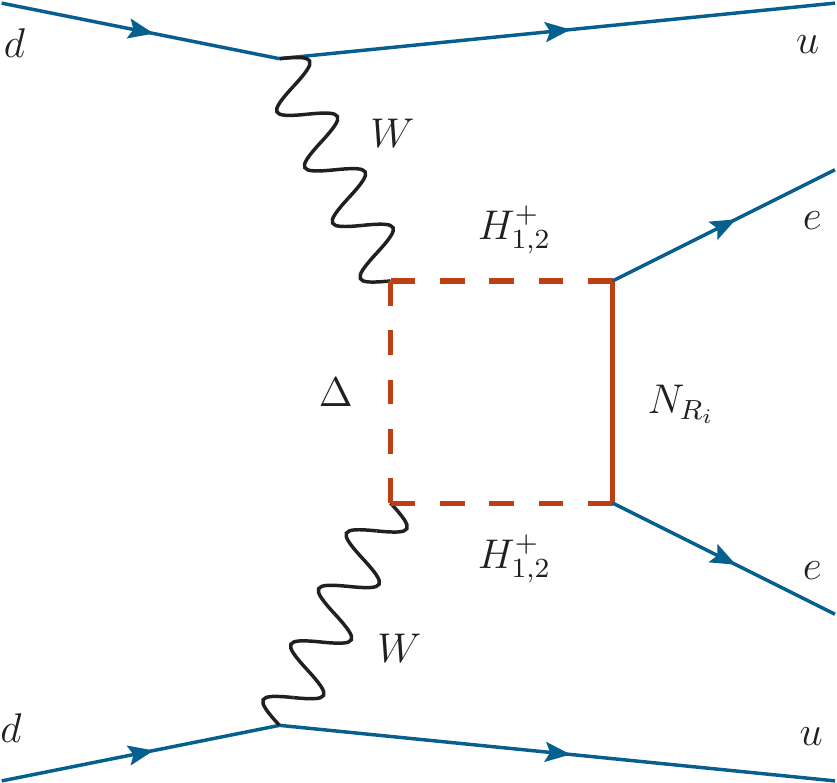}\hspace{0.5cm} 
\includegraphics[width=0.28\columnwidth]{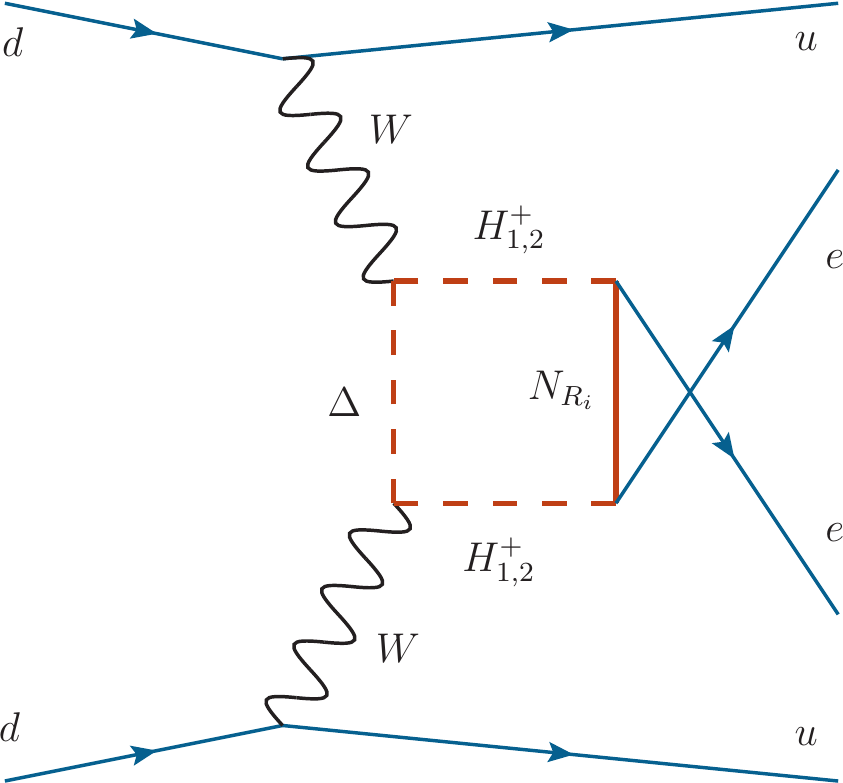}\hspace{0.5cm} 
\includegraphics[width=0.28\columnwidth]{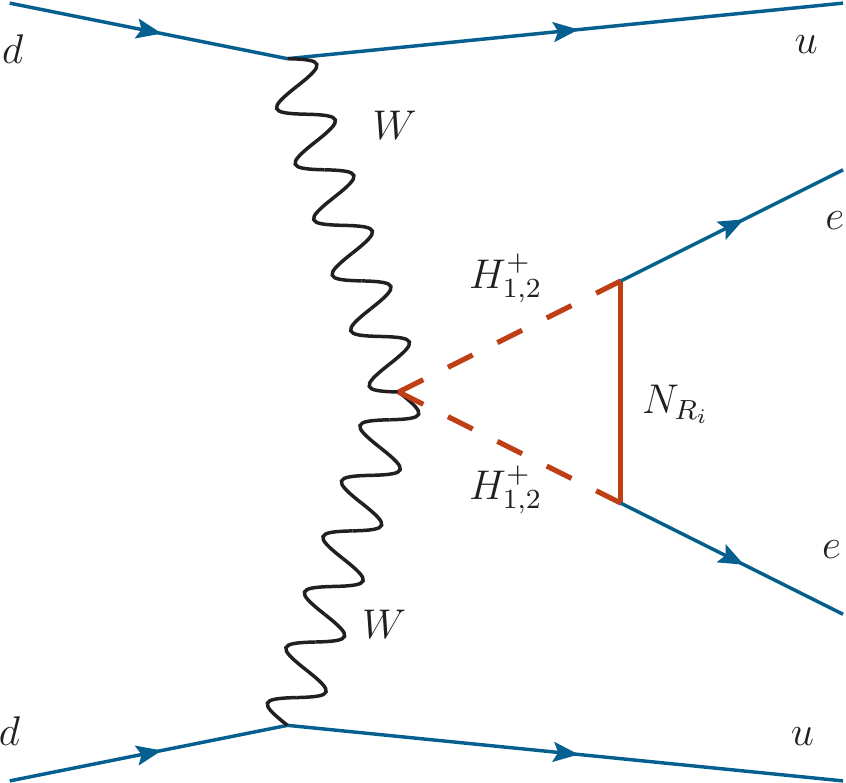}}
\caption{Neutrinoless double $\beta$-decay 1-loop contributions to $\mathcal{A}^{\mathrm{SD}}_{0\nu\beta\beta}$ for the {\sl Class 2} model from section~\ref{sec:class2_chinesemodel}. BSM propagators are depicted in red.}
\label{fig:App_D_Nu0bb}
\end{figure}

In addition, there is a further difficulty in simultaneously fitting the neutrino oscillation data and satisfying the constraints from neutrinoless double $\beta$-decay. The model from section~\ref{sec:class2_chinesemodel} 
generates 1-loop contributions to the short-distance neutrinoless double $\beta$-decay amplitude $\mathcal{A}^{\mathrm{SD}}_{0\nu\beta\beta}$, as shown in  Figure~\ref{fig:App_D_Nu0bb} 
and given by
\begin{equation}
\mathcal{A}^{\mathrm{SD}}_{0\nu\beta\beta} = \lambda_4^2 \times \frac{1}{(16 \pi^2)} \,\sum_{i=1}^{n}\, g_{ie}g_{ie}\, m_{N_i} \,\sum_{j=1}^{3}
\,K_j(m_{N_i}) = \frac{\lambda_4^2}{(16 \pi^2)} \,\sum_{i}\, g_{ie}g_{ie}\, m_{N_i} \,K_i
\label{int10}\\
\end{equation}
where the $K_j(m_{N_i})$ are the separate 1-loop integral contributions related to the topologies shown in Figure~\ref{fig:App_D_Nu0bb}. These contributions are not suppressed by $(m^{\ell}_e/v)^2$, as opposed to $m^{\nu}_{ee}$, and for $g_{ie} \sim 1$ the resulting $\mathcal{A}^{\mathrm{SD}}_{0\nu\beta\beta}$ in fact violates current experimental bounds~\cite{Geng:2015coa}, unless a cancellation among different $g_{ie}g_{ie}$ terms in Eq.~\eqref{int10} is present. 

It is nevertheless clear that a cancellation in $\mathcal{A}^{\mathrm{SD}}_{0\nu\beta\beta}$ may be arranged without affecting the model predictions for neutrino masses and mixings. For example, by adding a pair of states $N_{R_i}$ ($i = 4,5$) to the model with $n = 3$, such that $N_{R_{4,5}}$ only couple to electrons (not to muons or $\tau$-leptons) and demand
\begin{eqnarray}
\label{math1}
\sum_{i=4}^{5}\, g_{ie}^2 m_{N_i} K_i = - \sum_{i=1}^{3}\, g_{ie}^2 m_{N_i} K_i\,, \\
\sum_{i=4}^{5}\, g_{ie}^2 m_{N_i} \,I_i = 0 \, ,
\label{math2}
\end{eqnarray}
which is a linear system of equations for $g_{4e}^2$ and $g_{5e}^2$ that always has a solution (as long as $K_4/I_4\neq K_5/I_5$). 
This leaves $m^{\nu}_{ee}$ (generated by the contributions from $N_{R_i}$, $i = 1,2,3$) unaffected while yielding a total $\mathcal{A}^{\mathrm{SD}}_{0\nu\beta\beta} = 0$. Still, the introduction of more $N_{R_i}$ states with tuned couplings only to cancel the contribution to $\mathcal{A}^{\mathrm{SD}}_{0\nu\beta\beta}$ seems very much {\sl ad hoc}, and altogether highlights the generic difficulty of these models to fit neutrino oscillation data and satisfy bounds from neutrinoless double $\beta$-decay, needing in both cases strong tuning among model parameters.

\section{\label{sec:Conclusions} Conclusions}

Among $\Delta L = 2$ higher-dimensional SM operators responsible for the generation of neutrino masses, those involving SM gauge fields have so far not been explored in detail in the literature, despite their interesting properties. These include a link between the presence of SM gauge fields and the chirality of the SM charged leptons in the operator, together with an automatic loop suppression of neutrino masses and a sizable contribution to the short-distance neutrinoless double $\beta$-decay amplitude, which generally  dominates over the light Majorana neutrino exchange. In this work we have studied in detail the leading of such $\Delta L = 2$ operators with two right-handed charged leptons, appearing at $D = 9$ and labelled $\mathcal{O}_9$ throughout the manuscript. Neutrino masses from this operator are first generated at 2-loop order and further suppressed by the SM charged lepton masses via $m^{\ell}_a/v$, thereby providing a natural explanation for their smallness compared to the electroweak scale.   

We have analysed the structure and properties of BSM renormalizable completions to the operator $\mathcal{O}_9$, finding that there are two possible classes of models for such completions. We have discussed the general features of these completions, highlighting in particular the connection between neutrino masses and DM in these classes of models, with the DM candidate being an integral part of radiative neutrino mass generation.
A general feature is that the leading contribution to neutrino masses appears at 3-loop order. For each class, we have provided examples of specific radiative neutrino mass models (several of them genuinely new), and have discussed the various phenomenological aspects of these models, such as lepton flavour violation, collider signatures, the impact on electroweak precision observables and the DM properties. 

Finally, we have paid special attention to the interplay between neutrino mixing and neutrinoless double $\beta$-decay in these scenarios. 
Both for~{\sl Class~1} and {\sl Class 2} the dominant contribution to the $\Delta L = 2$ neutrinoless double $\beta$-decay process comes from the short-distance amplitude, rather than from the neutrino mass mechanism (light Majorana neutrino exchange). For {\sl Class 2} models, the structure of the neutrino mass matrix imposes strong cancellations among parameters in order to fit the neutrino oscillation data, and further cancellations in the neutrinoless double $\beta$-decay amplitude are needed to satisfy the current experimental bounds from e.g.~KamLAND-Zen (the possible viability of these models given the needed cancellations will be explored in \cite{GR}). This is in contrast with {\sl Class 1} scenarios, for which no such cancellations are required. 
Altogether, neutrino mass models from $\mathcal{O}_9$ turn out to be very predictive as a result of the many different phenomenological aspects they are linked to, which also result in these models (particularly for {\sl Class~2}) being severely constrained experimentally.

\acknowledgments
\noindent We thanks Diego Aristizabal for many useful discussions and comments on the manuscript. 
We also want to thank Arcadi Santamaria for discussions on radiative neutrino masses and model building, as well as 
Raymond Volkas for correspondence many years ago which led us to think more seriously about these issues. M.G.\ acknowledges partial support from 
the European Unions Horizon 2020 research and innovation program under grant agreement No 690575 and No 674896. M.A.R acknowledges partial support 
from Anillo grant PIA/ACT1406, CONICYT, Chile; Fondecyt grant 1171136, Chile and  Proyecto Interno USM PI.L.18.23 "Phenomenology Of Neutrinos and Dark Matter in 
Extensions of the Standard model". J.M.N. was supported by Ram\'on y Cajal Fellowship contract RYC-2017-22986, and also acknowledges support from the Spanish 
MINECO's ``Centro de Excelencia Severo Ochoa" Programme under grant SEV-2016-0597, from the European Union's Horizon 2020 research and innovation programme under 
the Marie Sklodowska-Curie grant agreements 690575  (RISE InvisiblesPlus) and 
674896 (ITN ELUSIVES) and from the Spanish Proyectos de I$+$D de Generaci\'on de Conocimiento via grant PGC2018-096646-A-I00.


\appendix

\section{Appendix: Oblique parameters in the inert triplet dark matter model}
\label{app:A}

In order to analyze the impact on the electroweak precision observables from the inert triplet model in section~\ref{sec:lollipop} (see also \cite{Alcaide:2017xoe}), we compute the oblique $S$, $T$ and $U$ 
parameters~\cite{Peskin:1991sw}. Following the notation of \cite{Maksymyk:1993zm}, they read
 \bea
 \frac{\alpha_{\mathrm{em}}}{4\,c_W^2 s_W^2}\, S & \equiv & \frac{\Pi_{ZZ}(m_Z^2)-\Pi_{ZZ}(0)}{m_Z^2}
 -\frac{c_W^2 - s_W^2}{c_W s_W}\Pi_{Z\gamma}^{'}(0)-
 \Pi_{\gamma \gamma}^{'}(0),\label{Sparam}\\
 \alpha_{\mathrm{em}}\, T & \equiv &  \frac{\Pi_{WW}(0)}{m_W^2} -\frac{\Pi_{ZZ}(0)}{m_Z^2}, \label{Tparam}\\
 \frac{\alpha_{\mathrm{em}}}{4 \,s_W^2} \, U & \equiv &  \frac{\Pi_{WW}(m_W^2)-\Pi_{WW}(0)}{m_W^2}
 -c^2_W \Big[\frac{\Pi_{ZZ}(m_Z^2)-\Pi_{ZZ}(0)}{m_Z^2}\Big] \nonumber \\ 
 &  &  -s^2_W\, \Pi_{\gamma \gamma}(0) -2\,s_W c_W \,\Pi_{Z \gamma}(0), \label{Uparam}
 \eea
where $\Pi(p^2)$ are the contributions to the gauge bosons' self-energies from the BSM fields, at the (squared) energy scale $p^2$. $\alpha_{\mathrm{em}}$ (the fine-structure constant), $s_W = \sin \theta_W$ (the sinus of the Weinberg angle), $m_W$ and $m_Z$ take their experimental values as inferred within the SM. 

When fairly low mass states are present in a model, additional parameters $V$, $X$ and $W$ must be included to more precisely describe the impact of BSM physics on electroweak precision observables~\cite{Maksymyk:1993zm}. For example, $s^2_W/(s^2_W)_\text{SM}$ is as function of $S$, $T$ and $X$; the ratio of decay widths $\Gamma(Z \to \nu \bar{\nu})/\Gamma_\text{SM}(Z \to \nu \bar{\nu})$ is a function of $T$ and $V$; the decay width ratio $\Gamma(W\to \text{all})/\Gamma_\text{SM}(W\to \text{all})$ is a function of $S$, $T$, $U$ and $W$. In this work we concentrate only on the BSM contribution to the most relevant $S$, $T$ and $U$ parameters, whith ther SM values set to 0 at a top quark mass $m_{t}= 173$~GeV, higgs mass $m_h = 126$~GeV and $\alpha_{\mathrm{em}}$ evaluated at the $m_Z$ scale~\cite{Tanabashi:2018oca}.

In order to obtain the oblique correction induced by an inert $SU(2)_L$ triplet scalar with hypercharge $Y=1$, a real $SU(2)_L$ singlet scalar, and a doubly charged $SU(2)_L$ singlet scalar, we derive the needed Feynman rules and evaluate the BSM contributions to the gauge boson self-energies $\Pi(p^2)$. 

It turns out to be convenient to introduce the following function
\be
B_5(p^2,m_i^2,m_j^2)=4\,B_{22}(p^2,m_i^2,m_j^2)- A(m_i^2) - A(m_j^2),
\ee
where $A$ and $B_{22}$ are the Passarino-Veltman scalar functions given in \cite{Passarino:1978jh} (for a modern approach to evaluate these functions see, e.g.~\cite{Romao}, but note that this reference use an opposite sign convention for the scalar functions).
All the contributions to Eqs.~(\ref{Sparam}-\ref{Uparam}) can now be written in term of the BSM particle masses ($m_{\Delta^{++}}$, $m_{\Delta^+}$, $m_{A_0}$, $m_{S_1}$, $m_{S_2}$ and $\rho^{++}$) through the $B_5$ function, and the singlet-triplet mixing angle~$\alpha$.
The BSM contributions to the gauge bosons' self-energies can be expressed as follows:
\bea
\Pi_{\gamma \gamma}(p^2) & = & \frac{\alpha_{\mathrm{em}}}{4\pi} \,
\left[ 4\,B_5(p^2, m_{\Delta^{++}}^2, m_{\Delta^{++}}^2) + B_5(p^2,m_{\Delta^{+}}^2,m_{\Delta^{+}}^2) \right]  +  \frac{\alpha_{\mathrm{em}}}{\pi}\, B_5(p^2,m_{\rho}^2,m_{\rho}^2) \nonumber \\
\Pi_{Z \gamma}(p^2)&=& \frac{\alpha_{\mathrm{em}}}{4\pi}\frac{1}{s_{W}c_{W}}\left[
 2\,(c_{W}^2-s_{W}^2)\,B_5(p^2,m_{\Delta^{++}}^2,m_{\Delta^{++}}^2)-s_{W}^2\,B_5(p^2,m_{\Delta^{+}}^2,m_{\Delta^{+}}^2)\right]
 \nonumber \\
 &  &- \frac{\alpha_{\mathrm{em}}}{\pi} \frac{s_{W}}{c_{W}} B_{5}(p^2,m_{\rho}^2,m_{\rho}^2) \nonumber \\
\Pi_{Z Z}(p^2)&=&\frac{\alpha_{\mathrm{em}}}{4\pi}\frac{1}{s^2_{W}c^2_{W}}\left[
 (c_{W}^2-s_{W}^2)^2 \, B_5(p^2,m_{\Delta^{++}}^2,m_{\Delta^{++}}^2)-s_{W}^4\, B_5(p^2,m_{\Delta^{+}}^2,m_{\Delta^{+}}^2)\right.\nonumber    \\
& & + \left. {c_{\alpha}^2}B_5(p^2,{m_{S_1}^2},m_{A_0}^2) + {s_{\alpha}^2}B_5(p^2,{m_{S_2}^2},m_{A_0}^2)\right]+\frac{\alpha_{\mathrm{em}}}{\pi} \frac{s_{W}^2}{c_{W}^2} B_{5}(p^2,m_{\rho}^2,m_{\rho}^2) \nonumber  \\
\Pi_{W W}(p^2) &=& \frac{\alpha_{\mathrm{em}}}{4\pi}\frac{1}{s^2_{W}}\left[
B_5(p^2,m_{\Delta^{++}}^2,m_{\Delta^{+}}^2) +  \frac{1}{2}B_5(p^2,m_{\Delta^{+}}^2,m_{A_0}^2)\right.\nonumber \\
& & + \left.\frac{{c_{\alpha}^2}}{2}B_5(p^2,m_{\Delta^{+}}^2,{m_{S_1}^2})+
\frac{{s_{\alpha}^2}}{2}B_5(p^2,m_{\Delta^{+}}^2,{m_{S_2}^2})\right], 
\eea
where we recall that $s_\alpha = \sin \alpha$ sets the singlet (triplet) content of the $S_1$ ($S_2$) field. 
Additional analytical simplifications are possible, e.g.\ $B_5(0,m,m)=0$, and any remaining expressions and their derivatives are numerically evaluated with the {\sf FF} package \cite{vanOldenborgh:1990yc}.

\vspace{2mm}

In order to derive constraints on BSM model parameters from the $S$, $T$ and $U$ measurements we evaluate the goodness-of-fit with the chi-square
\be
\chi^2 = \sum_{x,y \in\{S,T,U\}} (x_{0}-x) V^{-1}_{xy}(y_{0}-y)
\ee
From the particle data group \cite{PZyla:2020} we take the experimental best-fit values to be $S_0= -0.01\pm0.10$, $T_0=0.03\pm0.12$, $U_0=0.02\pm0.11$ with the errors denoting the 1-sigma variances $\sigma_x$, and $V_{xy}$ is the covariance matrix (or the error matrix) determined by
\be
V_{xy} = \sigma_x \sigma_y \rho_{xy}
\ee
with the symmetric correlations matrix $\rho_{xy}$ given by the entries: $\rho_{xx}=1$, $\rho_{ST}=0.92$, $\rho_{SU}=-0.80$, $\rho_{TU}=-0.93$.
A p-value of 5\% is reached at $\chi^2 = 7.82$  (i.e.\ the 5\,\% quantile for a $\chi^2$ distribution with 3~degrees of freedom). The best-fit models have $\chi^2\simeq0$. For example, $\chi^2\simeq0.032$ at the point $\lambda_6=1$, $s_\alpha=0.111$, $m_{\rho}=1$~TeV, $m_{\Delta^{++}}=2.337$~TeV and $m_{\Delta^+}=2.363$~TeV (implying that $m_{S_2}=2.308$~TeV, $m_{A_0}=2.390$~TeV, $m_{S_1}=2.391$~TeV) with the contribution $\{\Delta S\simeq0.004$, $\Delta T\simeq0.05$, $\Delta U\simeq 0\}$ in addition to the SM's $\{S_\text{SM}=T_\text{SM}=U_\text{SM}=0\}$.
The regions in the $m_{\Delta^{++}}-m_{\Delta^{+}}$ plane with $\text{p-values} > 5\,\%$ are shown in Figure~\ref{Fig_STU_Triplet}, where we made a scan over $50\leq m_{\Delta^{++}}\leq 3000$~GeV, $50\leq m_{\Delta^{+}}\leq 3000$~GeV and $-1\leq \sin{\alpha}\leq1$ for three fixed $\lambda_6 = 0.1, 1$ and $4\pi$, and for a fixed $m_{\rho} = 1$~TeV.
We also constrained $S_2$ to have a positive mass and to be the lightest among the $\mathbb{Z}_2$-odd states, for it to be potentially viable DM candidate.

\bibliographystyle{JHEP}
\bibliography{Neutrino3loop}

\end{document}